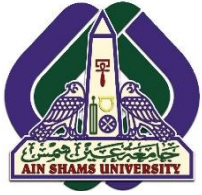

Scientific Computing Department
Faculty of Computer & Information Sciences
Ain Shams University

# High Performance Hyperspectral Image Classification using Graphics Processing Units

A thesis submitted in partial fulfillment of the requirements for the degree of
Master of Science
in
Computer and Information Sciences

By

## *Mahmoud Ahmed Hossam Edeen Mohammad*

B.Sc. in Computer and Information Sciences,
Teaching assistant at Basic Science Department
Faculty of Computer and Information Sciences
Ain Shams University

Under the Supervision of

### *Prof. Dr. Mohammad Fahmy Tolba*
Scientific Computing Department
Faculty of Computer and Information Sciences
Ain Shams University

### *Ass. Prof. Hala Muosher Ebied*
Scientific Computing Department
Faculty of Computer and Information Sciences
Ain Shams University

### *Dr. Mohammad Hassan Abdel Aziz*
Basic Sciences Department
Faculty of Computer and Information Sciences
Ain Shams University

Cairo 2015

# Acknowledgement

All praise and thanks to ALLAH, who provided me the ability to complete this work.

I am most grateful for my parents, who lovingly surrounded me with their care and overwhelming support to complete my studies.

I offer my sincerest gratitude to my supervisors. First and foremost, I would like to thank *Prof. Dr. Mohammad Fahmy Tolba* for his valuable guidance, support and motivation throughout the duration of this research.

I am greatly thankful to *Ass. Prof. Hala Muosher* for her meticulous efforts, patience and technical help throughout the research. I am equally thankful for *Dr. Mohammad Hassan* who helped me with his knowledge and experience.

I am deeply thankful for my family, specially my little sister for her sincere kindness and continuous support. I would like to specially thank my sincere friends *Ahmed Salah* and *Mahmoud Zidan* for their help, time and countless useful discussions. I am greatly thankful to my dear friend *Mohammad Magdy* for his sincere encouragement and technical advice in the last phase of the research. I thank all my wonderful friends and colleges who helped and supported me.

# Abstract

Real-time remote sensing applications like search and rescue missions, military target detection, environmental monitoring, hazard prevention and other time-critical applications require onboard real time processing capabilities or autonomous decision making. Some unmanned remote systems like satellites are physically remote from their operators, and all control of the spacecraft and data returned by the spacecraft must be transmitted over a wireless radio link. This link may not be available for extended periods when the satellite is out of line of sight of its ground station. In addition, providing adequate electrical power for these systems is a challenging task because of harsh conditions and high costs of production. Onboard processing addresses these challenges by processing data on-board prior to downlink, instead of storing and forwarding all captured images from onboard sensors to a control station, resulting in the reduction of communication bandwidth and simpler subsequent computations to be performed at ground stations. Therefore, lightweight, small size and low power consumption hardware is essential for onboard real time processing systems. With increasing dimensionality, size and resolution of recent hyperspectral imaging sensors, additional challenges are posed upon remote sensing processing systems and more capable computing architectures are needed. Graphical Processing Units (GPUs) emerged as promising architecture for light weight high performance computing that can address these computational requirements for onboard systems.

The goal of this study is to build high performance hyperspectral analysis solutions based on selected high accuracy analysis methods. These solutions are intended to help in the production of complete smart remote sensing systems with low power consumption. We propose accelerated parallel solutions for the well-known recursive hierarchical segmentation (RHSEG) clustering method, using GPUs, hybrid multicore CPU with a GPU and hybrid multi-core CPU/GPU clusters. RHSEG is a method developed by the National

Aeronautics and Space Administration (NASA), which is designed to provide more useful classification information with related objects and regions across a hierarchy of output levels. The proposed solutions are built using NVidia's compute device unified architecture (CUDA) and Microsoft C++ Accelerated Massive Parallelism (C++ AMP) and are tested using NVidia GeForce and Tesla hardware and Amazon Elastic Compute Cluster (EC2). The achieved speedups by parallel solutions compared to CPU sequential implementations are 21x for parallel single GPU and 240x for hybrid multi-node computer clusters with 16 computing nodes. The energy consumption is reduced to 74% using a single GPU compared to the equivalent parallel CPU cluster.

# Table of Contents

# List of Tables

# List of Figures

# List of Publications

# Chapter 1

# Introduction

# 1    Introduction

## 1.1    Real-Time Onboard Remote Sensing Systems

Remote sensing applications whether airborne or space borne provide huge benefits for important missions in a wide spectrum of fields ranging from scientific research, security and defense, agriculture, civil services, environmental studies and exploration. Some of these applications are time-critical and require real time or autonomous decision making, such as; search and rescue missions, target detection of military and defense deployment, risk or hazard prevention, wild land fire tracking, biological threat detection and monitoring of chemical contamination such as oil spills.

However, transmitting high-dimensional image data collected by airborne or satellite-based vehicle to a control station on Earth for processing may turn to be a very slow task, mainly due to the reduced bandwidth available and to the fact that the connection may be restricted to a short period. In the specific case of unmanned spacecraft systems (like satellites), the systems are physically remote from their operators, and all control of the spacecraft and data returned by the spacecraft must be transmitted over a wireless radio link as shown in Figure 1.1 . This radio link is low bandwidth, and may be unavailable for extended periods when the satellite is out of line of sight of its ground station, and the radio link often has high error rates. In addition, the high costs of production of these systems makes even providing the system with small power budgets a challenge [1].

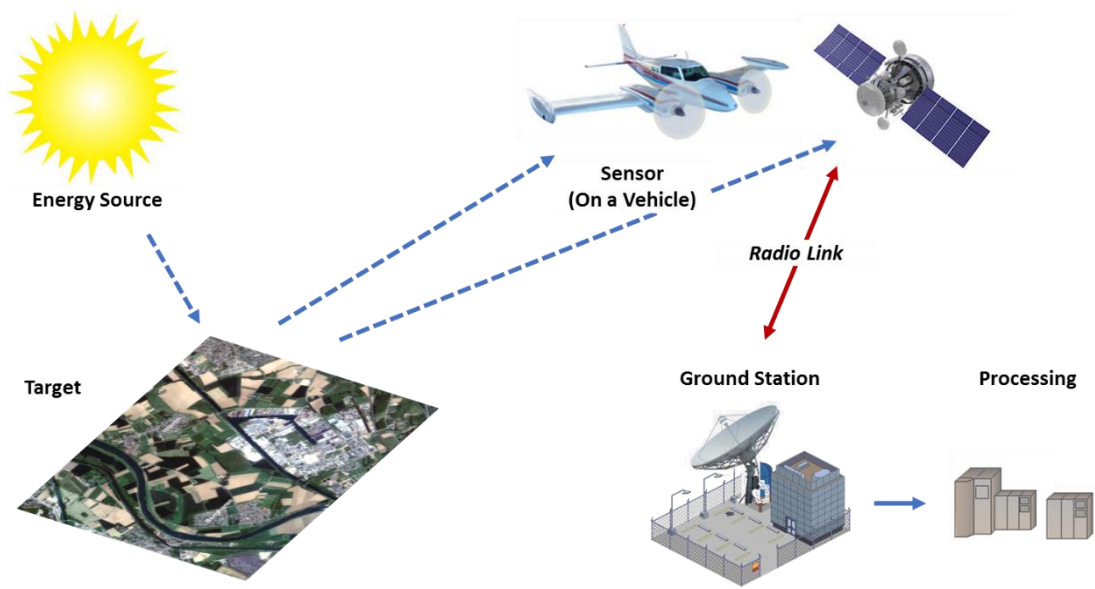

Figure 1.1. Components of remote sensing system, a remote sensor monitors a target and sends data to ground station for processing

Therefore, on-board processing is needed such that a significant portion of remote sensing data analysis is carried out on the vehicle, allowing for optional autonomous actions before sending data and feedback to the ground control station. The goal of the remote sensing mission is always towards smaller size, lower cost, flexible and high computational power onboard processing. Instead of storing and forwarding all captured images from onboard sensors, data processing can be performed on-board prior to downlink, resulting in the reduction of communication bandwidth and simpler subsequent computations to be performed at the ground stations [1]. An example of onboard processors is miniARCHER system from NovaSol [2], shown in Figure 1.2.

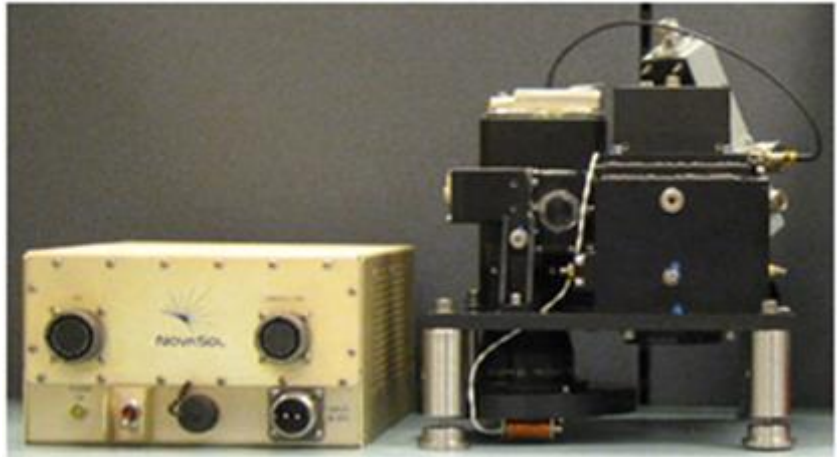

Figure 1.2. miniARCHER, a real-time onboard hyperspectral processing system from NovaSol. (Left: processor unit, Right: hyperspectral sensor)

A recent development in remote sensing is the introduction of hyperspectral imaging [3], in which images contain a large number, usually hundreds of wavelength bands, so that providing plenty of spectral information to identify spectrally unique materials as shown in Figure 1.3. Each pixel is a vector of light intensities or reflectance of sun light at different light wavelengths. A single pixel vector can be associated with one surface material of with a set of mixed materials with appropriate weights. For example, a pixel can be considered water only, or can represent both water and soil if the image resolution is small and pixels covers more land area with more than one material. The image analysis algorithms can benefit from the wealth of spatial and spectral information to more accurate analysis of remote sensing images. In turn, this wealth of data posed new challenges of high dimensional data and intensive time consuming computations. These high computational requirements, plus the fact that these systems will continue increasing their spatial and spectral resolutions, derived the researchers to investigate powerful computing platforms, which can efficiently handle high computational demands.

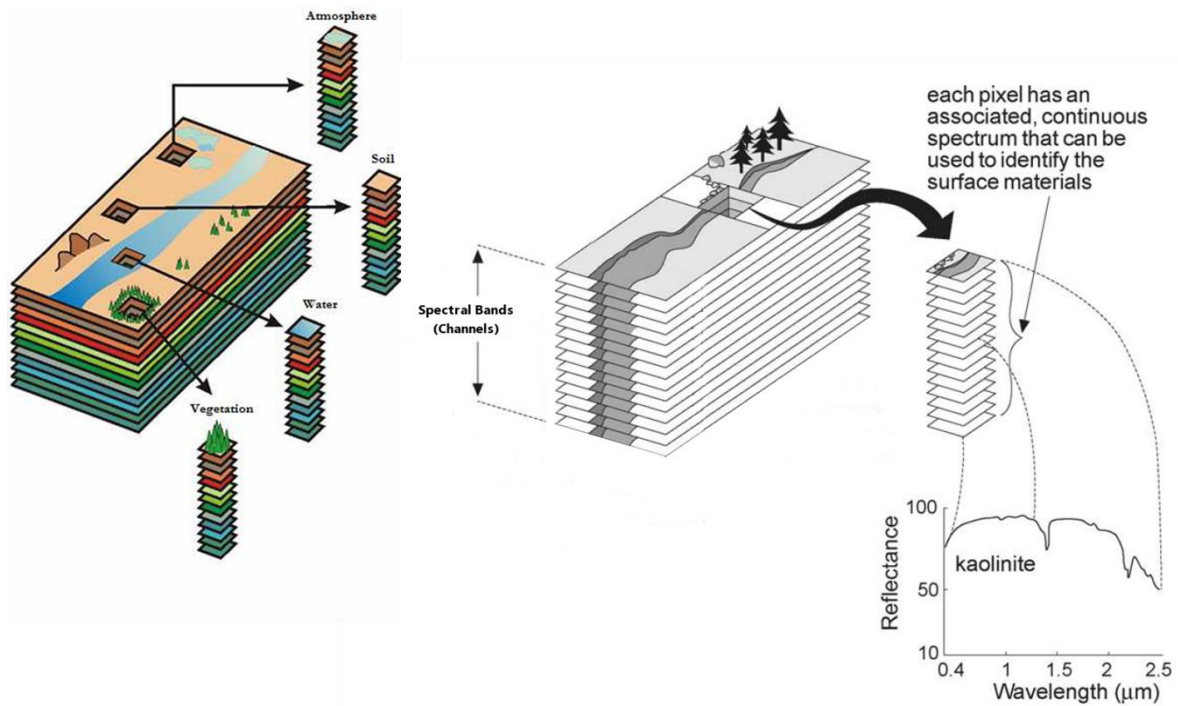

Figure 1.3. Hyperspectral image. A multi-channel image cube with each pixel vector represents a class of a certain material. The corresponding laboratory-measured spectral signature of the material is graph between material light reflectance and corresponding light wavelength.

High performance computing (HPC) is a suitable solution for such analysis systems that use high dimensional and large complex data like hyperspectral images. HPC can be achieved using high clock speed sequential processors or by using parallel computing platforms. However, the manufacturers of CPU chips are now faced by the clock wall of processing units, and the researchers found that the future of high performance computing depends on parallel and distributed computing rather than increasing the clock speed of single processing units [4].

GPUs has emerged recently as a promising platform for high performance computing. It captured the attention of researchers in a lot of research areas [5]. The computational power in GFLOPS (Giga Floating Point Operation per Second) of GPUs has grown much faster than the CPUs power over the last decade as shown in Figure 1.4. The important benefit besides the computational power of GPUs is the small size, lightweight and low power consumption. These make the GPUs a highly desired platform for remote sensing applications like satellite imaging and aerial reconnaissance [6]. Many hyperspectral analysis techniques have been implemented on parallel platforms, either on computer clusters or GPUs as discussed later.

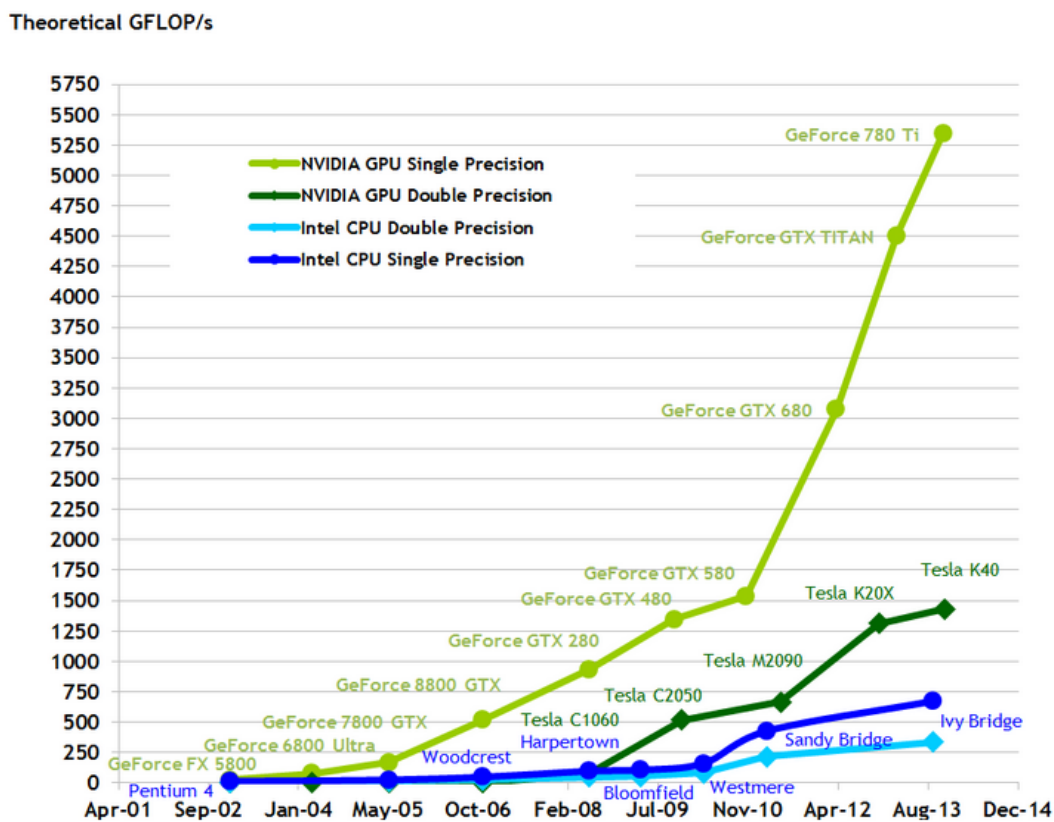

Figure 1.4. The evolution of computational power of GPUs measured in GFLOPs against CPU

## 1.2 Hyperspectral Analysis Methods

Remote sensing digital image analysis is a rich and vast research field that contains many different pattern recognition and statistical analysis methods. The choice of suitable analysis methods for a certain task is largely dependent on the nature of desired scenario and domain. Hyperspectral analysis methods can use the spectral information only or both the spatial and the spectral information of the image. The spectral methods treat pixel values as individual unrelated sets of spectral intensities with no particular spatial arrangement. The spatial-spectral methods take into account the pixel arrangement and the contextual entities in the image. The research of hyperspectral image analysis is increasingly moving towards spatial-spectral methods because of the importance of incorporating spatial and spectral aspects of data simultaneously, which has been recognized by many researchers in the field [7], [8], [9].

Hyperspectral analysis methods can be grouped under three main approaches [7]: Per-pixel analysis, mixed-pixel analysis and object-based image analysis (OBIA) [10]. Of these major approaches, there exist many classification/segmentation algorithms [11], [12] and spectral mixture analysis algorithms [13]. With the increase of spatial resolutions of new sensors, object-based image analysis (OBIA) has emerged as a promising approach to image analysis due to its efficiency with high spatial resolution images and the production of useful information about image classes and objects.

In the light of mentioned findings, it is desired to focus on spatial-spectral and object-based image analysis methods for better and more useful analysis and classification results. In addition, unsupervised analysis is encouraged because of the limited training samples, the difficulty of obtaining ground truth data in remote sensing and the need for automated responses for onboard processing [12]. Therefore, this work is concerned with unsupervised classification and clustering/segmentation approaches with spatial-spectral and object-based analysis. Recursive hierarchical segmentation (RHSEG) [14] is a well-known hyperspectral spatial-spectral OBIA method developed and used by the National Aeronautics and Space Administration (NASA). RHSEG has two main advantages: (1) Provides more accurate regions boundaries, and (2) Production of a hierarchical set of image segmentations with different detail levels from coarse to fine grain. However, hierarchical clustering methods are computationally intensive, especially when used with high dimensional data. In order to meet these computational challenges and provide solutions for onboard processing, suitable parallel solutions are needed.

## 1.3   Work Objective

The objective of this work is to build high performance hyperspectral analysis solutions based on selected high accuracy analysis methods using parallel and distributed GPUs architectures. These solutions are intended to help in production of complete smart remote sensing systems with low power consumption.

This objective is motivated by the emergence of hyperspectral imaging and Graphics Processing Units. Hyperspectral imaging has the potential of highly accurate analysis. GPUs provide desired architecture for low power high performance computing. The desired solutions can help build smart remote sensing systems like:

- Space exploration probes
- Autonomous unmanned airborne vehicles
- Deep underwater rescue systems

In this work, a parallel/distributed implementation of (RHSEG) is presented using GPUs, multicore CPUs, CPU clusters and hybrid multicore CPU/GPUs clusters, where shared memory architecture and distributed memory architectures are combined cooperatively and seamlessly. The speedup results are compared with sequential single CPU core, single multicore CPU and cluster of multicore CPUs. The fundamental idea of parallelizing and accelerating RHSEG is to distribute the most intensive dissimilarity calculation part among GPU threads and to partition the input image into sections, sending each section into multi-core CPU threads and cluster computing nodes. The GPU platforms that used for the proposed solution was NVidia's Compute Device Unified Architecture (CUDA) [15] and Microsoft C++ Accelerated Massive Parallelism (C++ AMP) [16]. The software platforms that are used for multi-core CPUs and distributed clusters are the QtConcurrent and the QtNetwork libraries by Digia [17], the proposed cluster solution is built using Amazon Elastic Compute Cloud cluster (EC2) [18]. Finally, Power and energy consumption for proposed solutions are investigated and compared against sequential and parallel CPU solutions.

## 1.4   Thesis Structure

The remainder of the thesis is organized as follows; Chapter 2 provides a detailed background on hyperspectral analysis methods and the reasons for selecting RHSEG method.

In chapter 3, we describe the existing high performance development platforms. Besides, we provide a background on the existing high performance implementations of hyperspectral methods.

Chapter 4 explains in detail the RHSEG method and presents proposed GPU/CPU solutions of RHSEG method: RHSEG for single GPU, hybrid CPU/GPU RHSEG and cluster CPU/GPU RHSEG.

Chapter 5 shows the obtained experimental results for different hyperspectral images. Finally, Chapter 6 concludes the results and mentions suggested future work.

# Chapter 2

# Literature Review

# 2 Literature Review

## 2.1 Introduction

Remote sensing image analysis as a subset digital image analysis is a rich and vast research field that contains many different methods and approaches for data analysis. With the emergence of hyperspectral sensors, new methods were introduced to the literature as well. There are different methods and techniques for hyperspectral analysis in almost all analysis phases; preprocessing, dimensional reduction, clustering, feature extraction and classification. Machine learning and image processing techniques have been applied to extract information from hyperspectral data in [19] and [20]. In addition, taxonomies and classifications of hyperspectral analysis methods and remote sensing algorithms have been developed by many researchers in the literature [13], [21] and [22].

## 2.2 Hyperspectral Analysis Methods

Hyperspectral analysis methods are categorized in various categorization criteria. From the nature of classification output point of view, methods can be grouped under three main approaches [7]: per-pixel analysis, mixed-pixel analysis and object-based image analysis. As described earlier in chapter 1, the physical model of sun light reflected from material surface used for analyzing remote sensing data is either considered represent a single material or group of materials. Per-pixel methods are classification or clustering methods that produce classification maps with each pixel assigned to only 1 class, meaning that reflected light intensities in the image represents only a single material. Mixed-Pixel methods produce classification maps with each pixel assigned to multiple of materials in different ratios or weights. There are many methods for these approaches, either for classification/segmentation algorithms [7], [23], [12] or spectral mixture analysis algorithms [21].

Other categorizations exist as well, from pixel arrangement point of view, analysis methods can be considered spectral methods or spatial/spectral methods [24]. Spectral methods processes input image as unordered set of pixels with no particular spatial arrangement, which means that the pixel position inside the image is irrelevant. If the pixel's position in the image is changed it will not affect the final classification of this pixel. On the other hand the spatial/spectral methods take the pixel spatial position in the image into consideration. The importance of incorporating spatial and spectral aspects of data simultaneously has been recognized by many researchers in the field [7], [25], and it is generally found that the use of contextual (or spatial) information provide better classification accuracies. For instance, urban area mapping requires sufficient spatial resolution to distinguish small spectral classes, such as trees in a park, or cars on a street [26], [27].

From the training samples point of view, analysis methods can be categorized as either supervised or unsupervised methods. In supervised methods, the human expert labeled training samples are used to extract features of desired classes, and then test images are classified based on the extracted trained features. Unsupervised methods process input images without the need for training samples. For remote sensing field, the small number of training samples and ground truth data and the high number of spectral features available in hyperspectral remote sensing data poses a challenge for analysis methods, as classification accuracy tends to decrease as the number of features increase [28], this is known as the Hughes effect [29]. Besides, the nature of on-board processing and absence of human experts and the need for automated responses makes the use of unsupervised methods more needed.

For all widely different analysis methods under different categories that exist in the literature, there is no clear comparative metric for deciding what methods are the best when used in different problem domains. Therefore, the choice of analysis method for a certain task is largely dependent on the nature of needed analysis and problem domain. In this work, we are concerned with per-pixel unsupervised clustering/classification approaches, which incorporate spatial/spectral features. This choice is urged by the limited training samples in the literature, need for automated on-board decisions and high accuracy results. Mixed-pixel classification is considered for future work.

## 2.3 Hyperspectral Image Per-Pixel Segmentation and Classification Methods

There are many different methods in the literature for per-pixel unsupervised hyperspectral image classification based on clustering or segmentation. These methods can be categorized under several main approaches [23]; Partitional clustering, Watershed transformation for segmentation, Graph methods for segmentation and hierarchical clustering. Partitional clustering is a classical approach, which is based on dividing the input image to arbitrary clusters and iteratively assigning the data points to these clusters using an error criterion measurement like the squared error.

The Watershed transformation for segmentation [30] uses the watershed contours that are generated from the input image as a boundary map for the segmentation process. The input image is considered as a topographic height map of pixels of intensity values, and the output watershed image represents the high boundaries around low points (or local minima areas) of the height map. The Watershed algorithm is originally calculated for gray scale single band images, but in [31], [32] it was adapted for multichannel images. Graph clustering methods [33] represent the image as a weighted undirected graph, where the pixels or the groups of pixels are the graph nodes and the weighted edges are the dissimilarity between adjacent pixels. After that, the graph is partitioned into smaller sub-graphs or trees that represent separate clusters. The partitioning process is carried out based on different criteria, like deleting edges with largest dissimilarity. Finally, the hierarchical clustering methods starts by assigning clusters to individual pixels, then merges these pixels iteratively based on similarity measure until desired number of clusters is reached. This approach generates multiple levels of classifications from fine grain close to individual pixels, to coarse grain at the final clusters.

Many methods exist in the literature for each of these four classification approaches. The segmentation techniques can be grouped into three classes working in the spatial domain, spectral domain or combining spatial-spectral domain [34]. A well-known example of partitional clustering is the Iterative Self-Organizing Data Analysis Algorithm (ISODATA) [35]. ISODATA is a spectral clustering method that use squared error criterion for clustering image pixels and does not incorporate pixel spatial information. The advantage of ISODATA is the low computational complexity, however these methods are sensitive to initial clusters generation method. The Watershed algorithm was used as a pre-segmentation step for enhancing the classification output [34]. In [36], Plaza developed unsupervised image classification/segmentation methodology by extending the watershed transformation to hyperspectral image processing. He compared this technique to a standard hyperspectral unsupervised classification algorithm; the ISODATA algorithm. Watershed transformation has low computation complexity compared to other segmentation techniques. However, Watershed transformation is known for over-segmentation of output regions and sensitivity to image noise [37]. Beucher [38] introduced a new algorithm called the waterfall algorithm to overcome the over-segmentation problem that usually comes with the watershed transformation.

An example of Graph clustering methods is the Minimum Spanning Forest (MSF) [39]. In this method, a graph *G* representing the initial classification of the image is generated by assigning each pixel to a graph vertex, and each edge connects couple of vertices and given a weight. This weight indicates the dissimilarity between these two vertices. A minimum spanning forest is then calculated using graph and the resulting sub-trees are used as regions for the final classification map. This method in combination with appropriate segmentation algorithms produces high accuracy results. However, depending on the initial classification method, if some regions are missed due to inappropriate classification parameters, these regions will be lost in the final classification map [40]. The Hierarchical Segmentation (HSEG) [41] is a hierarchical clustering method in which each pixel is considered as a separate region, and iteratively HSEG merges these regions until desired clusters number is reached. Each HSEG contains two possible merges; adjacent and non-adjacent regions merge. HSEG produces accurate region boundaries and high classification accuracy, but has high complexity and memory requirements. To address these challenges, a recursive approximation called Recursive HSEG (RHSEG) [14] was developed. Plaza et al. [12] used RHSEG clustering for unsupervised classification, which produced highly accurate classification results.

Few comparative studies have been conducted to compare analysis methods and techniques to each other in the literature. For instance, Fauvel et al. [34] studied and compared watershed, HSEG and MSF classifications for different hyperspectral datasets. HSEG based classification produced better overall classification results compared to watershed-based method in all datasets, the best overall accuracy in one of datasets compared to MSF based method, and the second best in the other dataset. A multiple classifier incorporating all three methods achieved the best overall accuracy for all datasets.

In Plaza et al. [12], a selected group of hyperspectral clustering/classification methods are studied in depth. The selected methods are determined by many considerations, such as the effectiveness with high-dimensional data, the incorporation of contextual or spatial information, their competitive classification accuracy compared with other analysis methods, the proposed remote sensing application domain and the availability of complete date with suitable ground-truth information. The methods studied included Contextual Support Vector Machine, Morphological Profiles, Markov Random Field-based (MRF) contextual classifier, automated morphological end-member extraction (AMEE) [25] and Recursive Hierarchical Segmentation (RHSEG). These methods were compared against other well-known remote sensing analysis methods such as standard (spectral) Support Vector Machines, Pixel purity index (PPI) [42] and others. It is finally found that the selected methods classification accuracy were more effective than other methods compared against in certain application domains. MRF contextual classifier produces high classification accuracy, but has the problem of incorrect region edges in the classification map output [43] [44] [45]. Table 2.1 shows the summery of investigated per-pixel analysis methods, HSEG and its approximation RHSEG are unsupervised methods that can produce highly accurate results with accurate region edges, but are very computationally expensive as described later in chapter 3. This work focuses on accelerating RHSEG using parallelization on distributed clusters, multicore CPUs and GPUs. In Figure 2.1 classification maps of hyperspectral image is shown for each investigated method, RHSEG classification map produces highly correct regions and edges with different image details and with less parameters tuning required.

Table 2.1. Summary of investigated per-pixel hypersepctral analysis methods

| Method Name | Spatial/Spectral integration | Method Nature | Advantages / Disadvantages |
|---|---|---|---|
| **Standard SVM** | No | Supervised | Reduces sensitivity to Hughes effect, but produces noisy output |
| **Contextual SVM** | Yes | Supervised | Produces more accurate results at region edges than standard SVM |
| **ISODATA** | No | Unsupervised | Computationally less complex, but sensitive to initial class generation |
| **Feature Extraction + Minimum Spanning Forest (MSF)** | Yes | Supervised | Produces highly accurate results, but sensitive to underlying classification or segmentation methods used |

| Feature Extraction / SVM + Markov Random fields (MRF) | Yes | Unsupervised / Supervised | Produces highly accurate results, but requires parameters tuning for good results. |
|---|---|---|---|
| Watershed-based Methods | Yes | Unsupervised / Supervised | Computationally less complex, but produces over segmented output |
| Hierarchical Segmentation (RHSEG) | Yes | Unsupervised / Supervised | Produces high accurate results and very accurate region edges, but very computationally expensive |

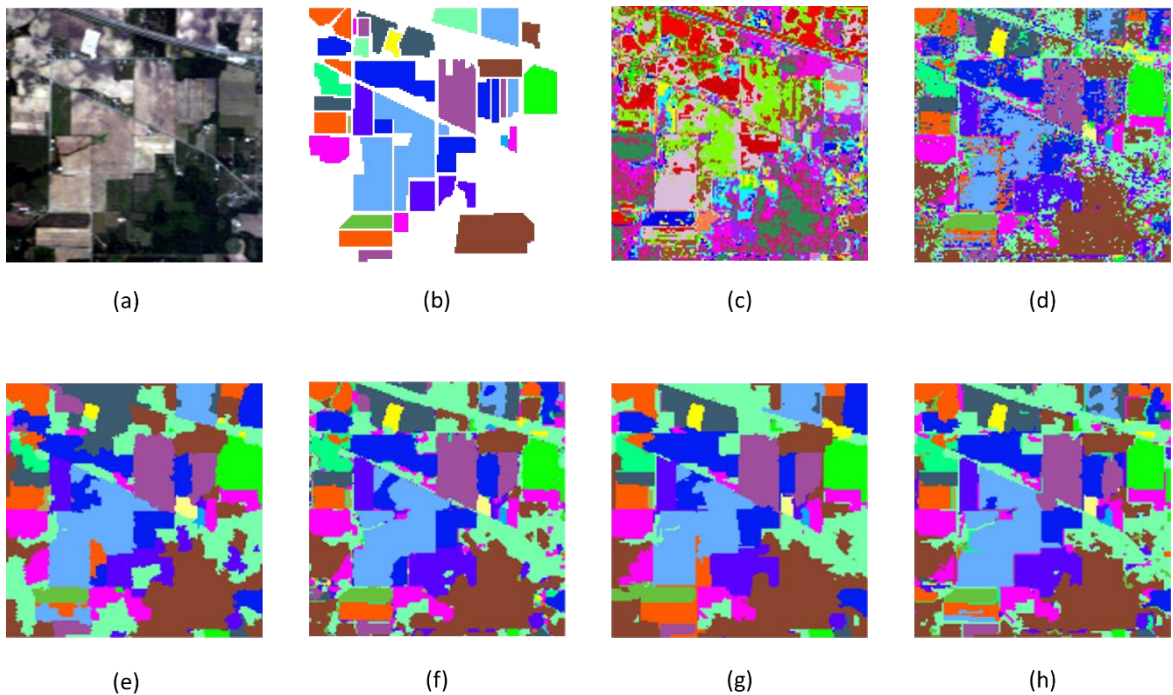

Figure 2.1. Classification maps examples of investigated analysis methods. (a) Hyperspectral image, (b) 16 class ground truth, (c) ISODATA, (d) SVM, (e) Feature Extraction + Watershed Segmentation, (f) Feature Extraction + MRF, (g) Feature Extraction + MSF, and (h) Feature Extraction + RHSEG

# Chapter 3

# Overview of Parallel Computing

# 3 Overview of Parallel Computing

## 3.1 Introduction

Parallel high performance computing architectures are divided into different models according to the relation between instruction and data in the execution; Single Instruction-Multiple Data (SIMD), Multiple Instruction – Single Data (MISD), Multiple Instructions – Multiple Data (MIMD). The most widely existing architectures are SIMD and MIMD. SIMD distributes multiple data over one instruction at a time in parallel, while MIMD is more flexible by launching multiple instructions with different data at the same time in parallel.

Graphics Processing Units (GPUs) are a special type of SIMD architectures called Single Instruction – Multiple Threads (SIMT) as in Figure 3.1. SIMT is a combination of SIMD and SPMD (Single Program Multiple Data), while multicore CPUs are well-known MIMD architectures. Programming for GPUs requires choosing a development platform out of many existing software and hardware platforms. For example, GPU manufacturers like NVidia and AMD/ATI provide different devices that can be used for parallel computation, either standalone graphics boards or as mobile/embedded graphics chips. There are several software platforms that can be used with these hardware devices, for example; NVidia provides its own Compute Device Unified Architecture (CUDA) platform that can run only on NVidia devices. Khronos group developed a similar platform called OpenCL [46] that can run on any supported device from any manufacturer. Recently Microsoft also developed its GPU/Multicore CPU software platform; C++ AMP (Accelerated Massive Parallelism) which can run on any graphics device from any manufacturer that support DirectX [47] technology.

The following subsections sheds a light on CUDA, C++ AMP and OpenCL platforms in some details. The main factors that were given the highest priority for selecting the desired platforms are the platform maturity and flexibility. Thus two of these platforms were selected; NVidia CUDA and Microsoft C++ AMP, those were the two most mature and advanced GPU platforms that also provide top computation performance.

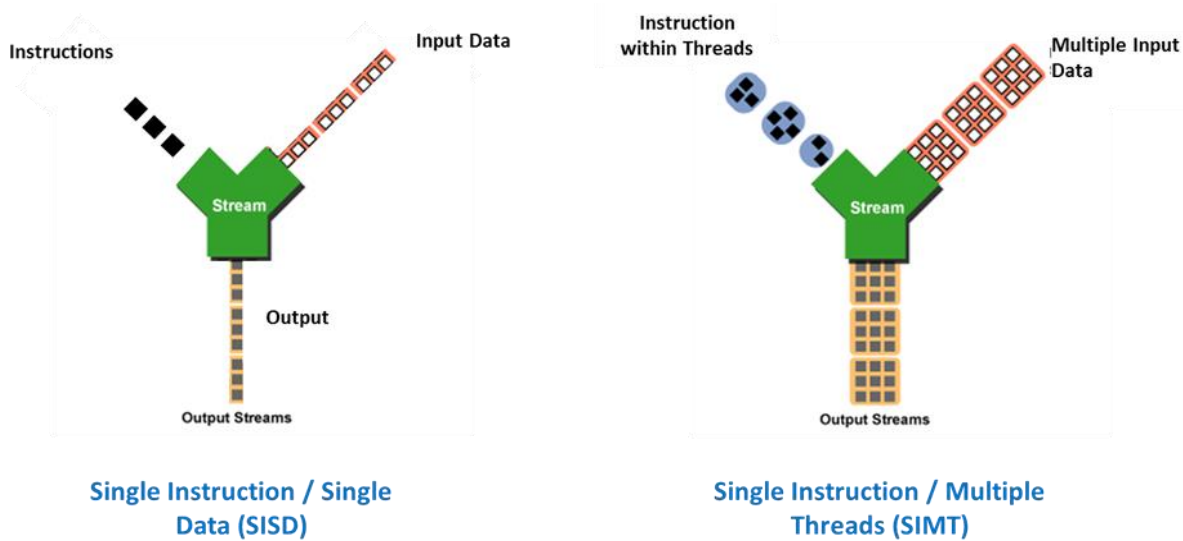

Figure 3.1. Illustration of SIMT, multiple input streams are processed by an instruction in multiple threads "kernels" at the same time in parallel.

## 3.2 GPU Platforms

### 3.2.1 NVidia Compute Unified Device Architecture (CUDA)

CUDA [15] is parallel shared memory architecture, in which the device is divided into multiple streaming multiprocessors (SMs), each multiprocessor has multiple of simpler streaming processors (SPs) as shown in Figure 3.2. All multiprocessors are connected to a cached high bandwidth global memory (up to and higher than 100 GB/s) using a network bus. In addition, there is low latency high bandwidth memory called shared memory, accessible by all SPs inside a multiprocessor. From the software point of view, there are three key concepts; threads inside hierarchal groups, shared memories, and barrier synchronization.

The parallel program is partitioned into coarse sub-problems that can be solved independently in parallel by blocks of threads and each sub-problem into finer pieces that can be solved cooperatively in parallel by all threads within the block.

This decomposition enables automatic scalability. Each block of threads can be scheduled on any of the available multiprocessors within a GPU, in any order, concurrently or sequentially, so that a compiled CUDA program can execute on any number of multiprocessors as illustrated in Figure 3.3, and only the runtime system needs to know the physical multiprocessor count. Thread blocks are executed independently; it can be executed in any order, in parallel or serially.

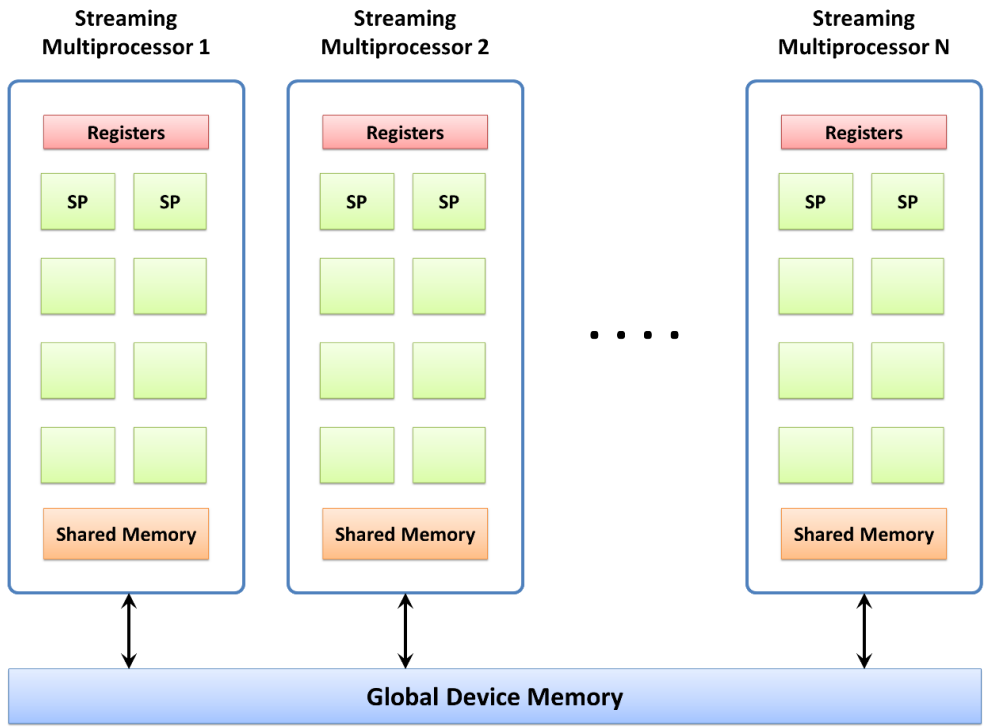

Figure 3.2. CUDA hardware architecture

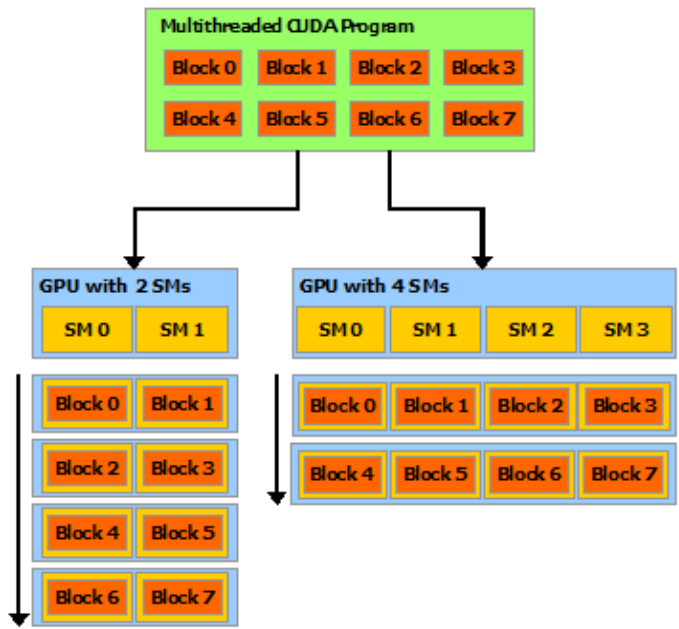

Figure 3.3. Automatic scalability of CUDA program execution, a multithreaded program is partitioned into blocks of threads that execute independently from each other, so that a GPU with more streaming multiprocessors (SM) will automatically execute the program in less time than a GPU with fewer multiprocessors.

Number of threads per block has a limit, since all threads of a block are expected to reside on the same processor core and must share the limited memory resources of that core. Blocks can be organized into a one-dimensional, two-dimensional, or three-dimensional grid of thread blocks as shown Figure 3.4. It is encouraged to launch as many threads as possible, even it exceeds the number of processors on the device, so that multiprocessors are kept busy most of the time, therefore making best use of the device capability.

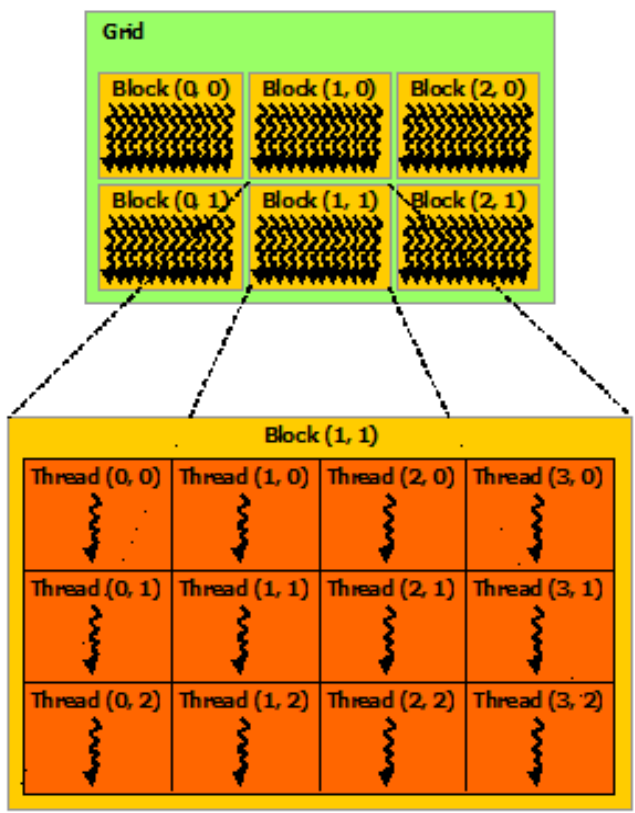

Figure 3.4. Grid of two-dimensional thread blocks

Threads within a block can cooperate by sharing data through some shared memory and by synchronizing their execution to coordinate memory accesses. Developers can specify synchronization points in the kernel by calling *syncthreads()* function, which acts as a barrier at which all threads in the block must wait before any is allowed to proceed.

**CUDA Memory Hierarchy**

From hardware point of view, there are two types of GPU memory, device memory and on-chip memory. Device memory is the main large memory of the device that is connected to all multiprocessors. Device memory is slower to access than the faster on-chip memory inside each multiprocessor. From software point of view, CUDA threads have access to data in many memory spaces during their execution as illustrated in Figure 3.5. Each thread has private local memory. Each thread block has shared memory visible to all threads of the block. All threads have access to the same global memory.

Two additional read-only memory spaces exist that are accessible by all threads: the constant and texture memory spaces. The global, constant, and texture memory spaces are persistent across kernel launches by the same application.

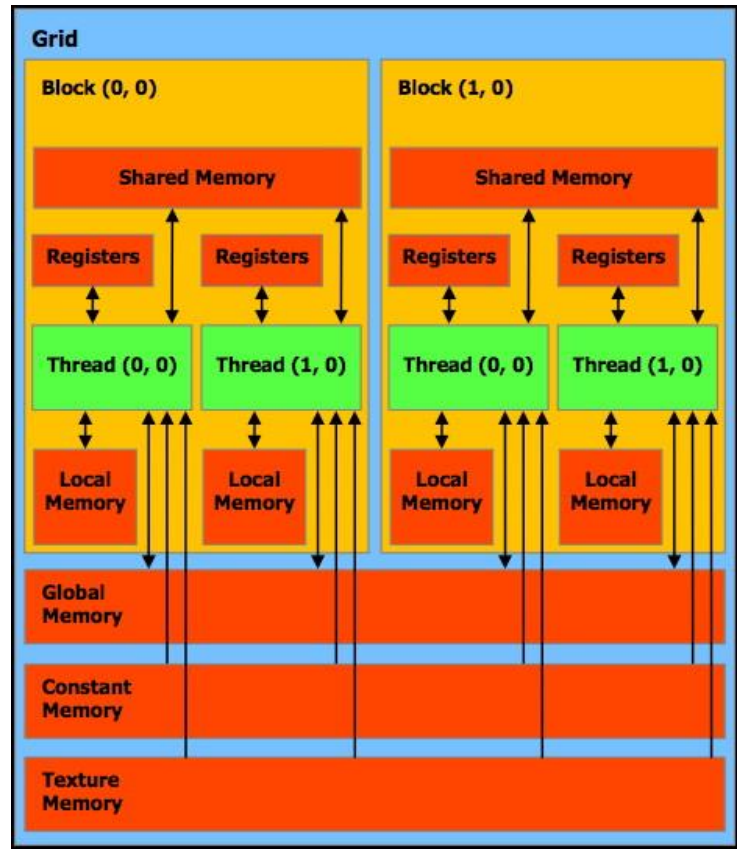

Figure 3.5. Different memory types in CUDA. All threads can communicate through global memory, and threads of the same block can communicate through the much faster shared memory.

Shared-memory space resides on the on-chip memory inside each multiprocessor while global, local, and constant and texture memory spaces reside on the slower device memory. To achieve high bandwidth, shared memory is divided into equally-sized memory modules, called banks, which can be accessed simultaneously. Any memory read or write request the made of $n$ addresses that fall in $n$ distinct memory banks which can therefore be serviced simultaneously. If two addresses of a memory request fall in the same memory bank, there is a bank conflict and the access has to be serialized.

To execute a parallel program on the GPU device, all relevant data structures and variables that reside on the host CPU memory have to be transferred to

GPU device. This memory transfer is done explicitly through CUDA memory APIs

### 3.2.2 Microsoft C++ AMP

Microsoft C++ AMP is an open-specification C++ language extension that enables developers to write parallel code for both multicore CPUs and GPUs. Its purpose is intended for simpler and cleaner parallel development than CUDA or OpenCL (C language based platforms). C++ AMP adheres to almost same software concepts of CUDA and OpenCL, but with much simpler framework especially for processing arrays across host memory (CPU memory) and GPU memory. C++ AMP is built on DirectCompute [48] technology, which is built on DirectX technology, therefore it can be run on any DirectX capable device from desktop devices to mobile chips.

### 3.2.3 Khronos OpenCL

OpenCL [46] is an open standard parallel programming framework for programming GPUs and multicore CPUs. It is widely supported by existing hardware from different manufacturers and shares the same software concepts with CUDA. By the time of this work, OpenCL was in its initial stages compared to CUDA which was already much mature and sound technology. OpenCL version of this work is considered in future as needed.

## 3.3 Other Parallel and Distributed Platforms

### 3.3.1 Multicore CPU Platforms

Multicore CPUs are processors with multiple identical computing cores and connected to one another through shared memory. Multicore CPU programming is MIMD in nature with large cache memory sizes suitable for different threads and instructions execution. There are many software platforms for multicore software development like OpenMP [49], Intel Threading Building Blocks (TBB), Microsoft Task Parallel Library (TPL) or Parallel Patterns Library (PPL) and Qt's QtConcurrent Library. The QtConcurrent library is used within this work and chosen for its compatibility with Qt framework that is already used for the development of our work. It should be noted that there are little differences between those platforms and most of them are suitable. They also can be used for the development of the proposed parallel multicore methods.

### 3.3.2 Computer Cluster Platforms

There are many hardware service providers that provide customized cluster computing service with paid fees. Examples of these services are Penguin Computing [50], Microsoft Azure, Google Cloud Compute Engine and Amazon Elastic Compute Cloud EC2. At the time of the development of this work, Penguin Computing and Amazon EC2 was the only services that provided GPU cluster computing. Amazon EC2 was used for its excellent support and flexible configurations. The used EC2 cluster is running on Windows Server operating system.

## 3.4    High Performance Implementations of Hyperspectral Methods

Hyperspectral clustering methods are computationally intensive due to high data dimensionality. Therefore, suitable parallel solutions are needed to overcome computational and memory requirement challenges. These methods have been implemented on various and different parallel architectures, parallel multi-processors, heterogeneous and homogeneous network of distributed computers and specialized hardware such as; field programmable gate arrays (FPGAs) and GPUs hardware architecture.

For example, ISODATA was parallelized using a Thunderhead CPU cluster [51] with 9x speedup using 16 processing nodes and also  parallelized using hybrid CPUs and GPUs [52] using Kenneland supercomputer [53] with hybrid nodes of Intel Xeon E5 8-core CPUs and NVidia M2090 GPUs. They achieved a speedup of 2.3x for distributed parallel GPU over distributed parallel CPU using 36 nodes. ISODATA was also parallelized on a single NVidia Kepler K20 GPU in [54] achieving 45x over sequential CPU implementation for 50 clusters of an output image. The Watershed based classification [36] was parallelized using a Thunderhead cluster achieving 13x speedup using 16 nodes and 170x speedup using 256 nodes.

RHSEG was parallelized using cluster CPUs and GPUs in [12] and [55] respectively. In [12] a homogenous Thunderhead Beowulf cluster at NASA's Goddard Space Flight Center is used to accelerate RSHEG. The Beowulf cluster [56] is composed of dual 2.4 GHz Intel Pentium 4 Xeon nodes, 256 GB DDR memory (1.0 GB of main memory available per CPU) and connected with a 2.2 GByte/s fiber interconnection system. The speedups achieved for these algorithms were 13x using 16 CPU nodes and 82x using 256 CPU nodes. In [55] GPU RHSEG is implemented using one-dimensional dissimilarity calculation kernel that processes every region dissimilarity calculations with all other regions in single thread per region. In addition, a hybrid multi-core CPU/GPU cluster was used for cooperative processing between CPU cores and the GPU for different image sections. Using a single NVidia GeForce 550 Ti board, an average speedup of 3.5x was achieved over sequential Intel Core i5 CPU implementation. With the use of hybrid 8-core Intel Xeon X5570 operating and NVidia Tesla M2050 GPU, additional average speedup up to 6x was achieved, using multicores cooperatively beside the GPU. Finally, using 16 node hybrid CPU/GPU clusters each having a single GPU resulted in a total of 112x speedup.

In the previously mentioned implementations of parallel RHSEG, either on GPUs or computer clusters, no energy consumption assessment was conducted. The aim of this thesis is to improve parallel implementations of RHSEG and to provide energy consumption assessment for the parallel solutions presented

# Chapter 4

# Proposed Parallel Methods

# 4    Proposed Parallel Methods

## 4.1    Recursive Hierarchical Segmentation (RHSEG) Method

As previously described in chapter 2, RHSEG method is an agglomerative hierarchical clustering method. The basic idea of hierarchical clustering in general is to start by assigning regions to individual pixels (or a small number of pixels), and then merge these regions iteratively based on the similarity measure until desired number of clusters is reached.    This approach generates multiple levels of classifications from fine grain close to individual pixels, to coarse grain at the final clusters. Figure 4.1 shows the concept of hierarchical segmentation on different levels of an image form starting level 1 (fine grained regions) to the final level 5 (final desired clusters). The image starts with six small regions, and by increasing the level number, the most similar regions are grouped (merged) into larger regions. At the final level 5, two clusters remain as the coarse grained classification result. The expert user is free to choose the desired classification result from output levels that best match the analysis needs.

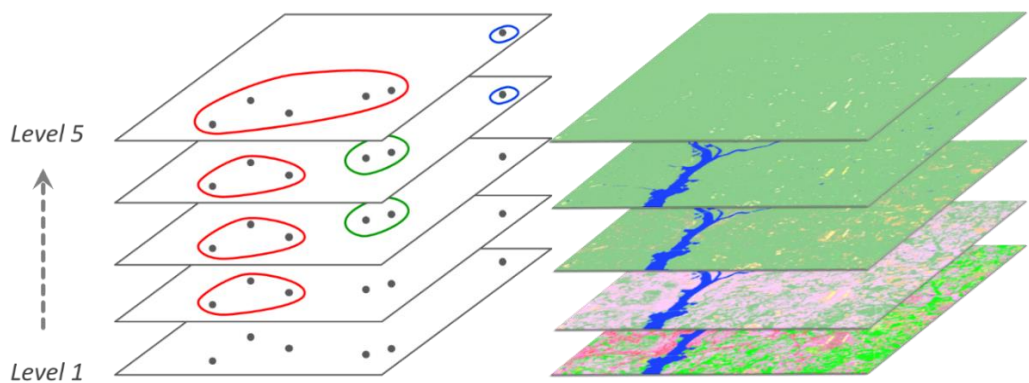

Figure 4.1. Concept of hierarchical clustering/segmentation by region growing. The lowest level has six initial regions, reduced by merging most similar regions with increasing clustering level until two coarsest clusters are reached at final level 5.

Recursive HSEG (RHSEG) described in this work is the recursive approximation of the computationally expensive Hierarchical segmentation (HSEG) method. Tilton [41] has developed HSEG method that is a combination of region growing and spectral clustering. HSEG method adds a new feature to the Hierarchical Step-Wise Optimal (HSWO) segmentation Algorithm [57] that is the addition of a spectral clustering step, which allows for the merge of non-adjacent regions controlled by the "spectral clustering weight" input parameter.

HSEG can be summarized in four steps:

1. Initialize the segmentation by assigning each image pixel a region label. If a pre-segmentation is provided, label each image pixel according to the pre-segmentation. Otherwise, label each image pixel as a separate region.

2. Calculate the dissimilarity value between all pairs of spatially adjacent regions, find the pair of spatially adjacent regions with the smallest dissimilarity value, and merge that pair of regions.

3. Calculate the dissimilarity value between all pairs of spatially non-adjacent regions, and find a pair with the smallest dissimilarity value, that is smaller than the minimum dissimilarity value found in (2). If found, then merge that pair of regions. If not, just go to step (4)

4. Stop if no more merges are required (min number of regions reached). Otherwise, return to step (2).

HSEG is an iterative region merging process, initialized with every pixel as a region. Figure 4.2 shows an outline of its main procedures, at each step, the dissimilarity value is calculated for each pair of spatially adjacent regions. The pair of regions with the smallest dissimilarity value is chosen for merging, and then the new merged region replaces them. Then the same step is repeated for non-adjacent regions. This process continues until the desired number of regions (segments or classes) is reached. The method can also be terminated automatically by checking a global convergence criteria to stop region growing iteration at certain threshold. HSEG is very computationally intensive, because it requires the calculation of the dissimilarity criterion value between each region and every other region in the image, which makes HSEG of order $O(N^6)$ in the worst case, where N is the edge length of input square image. Tilton [14] described the recursive implementation of this segmentation approach (RHSEG) on a cluster. Figure 4.3 shows a flowchart of divide-and-conquer, a recursive approach for implementing the HSEG method.

RSHEG approximates HESG by dividing the input image into $4^{(L-1)}$ sections, where *L* is the number of desired recursive levels for approximation. RHSEG starts at level 1 and divides the image until it reaches the deepest recursive level *L*, then it applies HSEG for each of the four image sections in the current deepest level. After that, the four sections in the deepest level are re-assembled back into larger image section, and HSEG is applied again on the reassembled image section. After RHSEG is finished with applying HSEG at the deepest levels L, the resulting four section images are reassembled back in to larger image of the previous recursive level and HSEG is applied to the reassembled image at the current level. This process of image reassembling and applying HSEG continues recursively from the deepest recursive level up to the first recursive level (Level 1). Small divided image sections are reassembled creating larger and larger images with HSEG applied on its sections till the final level is reached with the original image size fully clustered across recursive levels. Figure 4.4 shows how four image sections of any recursive level are reassembled. The regions along the edges of the four sections are linked together in an 8-neighborhood fashion, where each region in the edge registers the n-neighboring regions on the other edge as an adjacent region.

A wide variety of dissimilarity measure functions can be used in HSEG like Euclidean distance, vector norms (1-norm, 2-norm and infinity-norm), spectral angle mapper, spectral information divergence, mean squared error, normalized vector distance and image entropy. The choice of which dissimilarity function is better for the classification results, depends on the image domain and the type of analysis needed. For the experiments on urban and agricultural satellite images in this work the square root of band sum mean squared error (square root of BSMSE) produced better classification results than other functions and is used for the dissimilarity measurement which is given between any two regions $i$ and $j$ in an image of B bands by:

$$Square\ root\ of\ BSMSE(i,j) = \sqrt{\frac{n_i n_j}{(n_i + n_j)} \sum_{b=1}^{B} (\mu_{ib} - \mu_{jb})^2}$$

(1)

where $\mu_{ib}$ and $\mu_{jb}$ are the mean values for regions $i$ and $j$ in spectral band b, respectively. $n_i$ and $n_j$ are the number of pixels in regions $i$ and $j$ respectively.

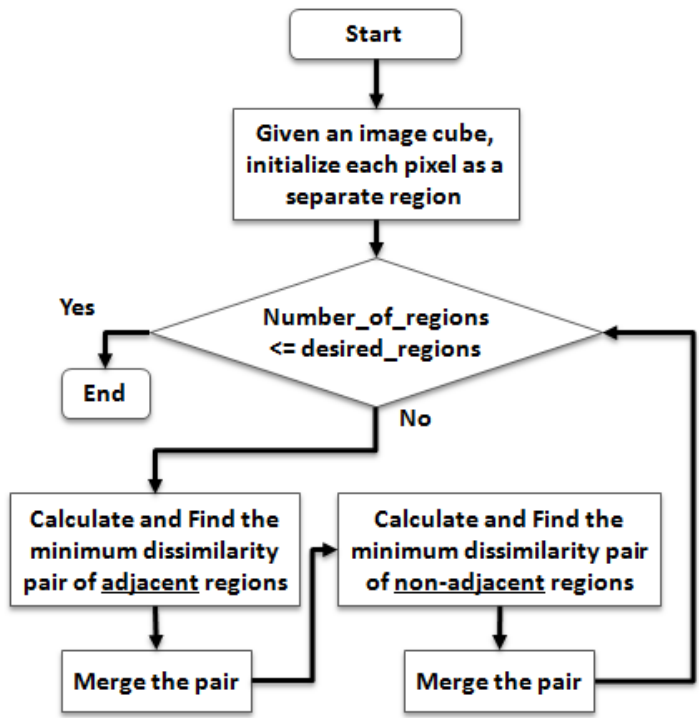

Figure 4.2. Outline of HSEG method.

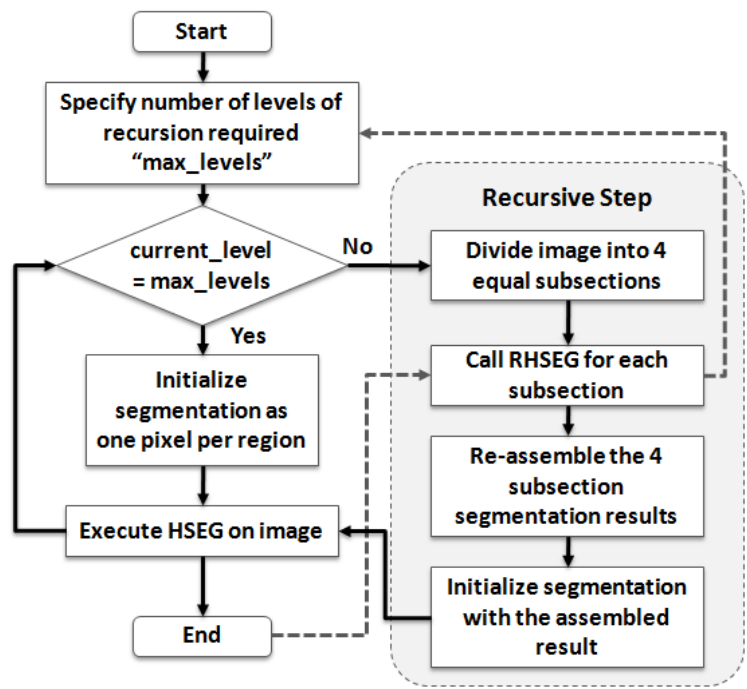

Figure 4.3.Flowchart of RHSEG method.

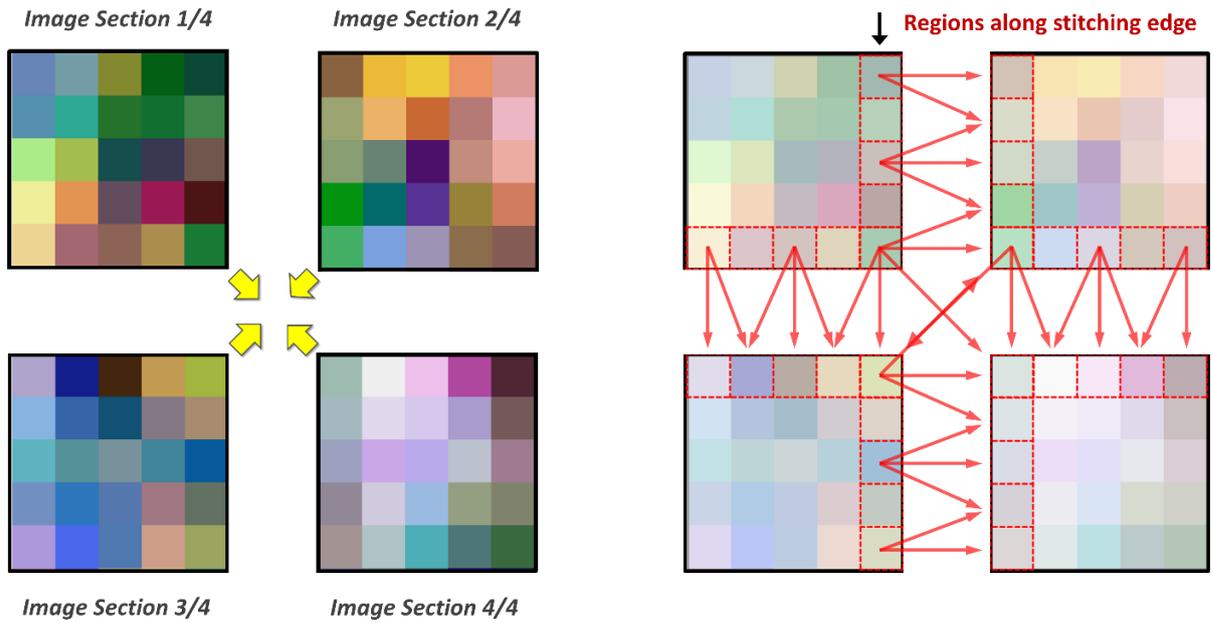

Figure 4.4. Reassembling of RHSEG image sections. Four image sections are re-assembled together into one image by linking regions along the edges with corresponding neighbor regions on the other side of the edge

## 4.2 Parallel Implementation of RHSEG on GPUs

This section presents a parallelization solution proposed for an RHSEG algorithm using GPUs. The main idea of parallelizing the RHSEG algorithm is to distribute the computation of a pair of regions for a dissimilarity measurement to the massive number of GPU threads in parallel. This is the most computationally intensive task of the whole algorithm and it takes over 95% of the whole execution time. We propose two different approaches to distribute the dissimilarity measurement between regions among GPU threads, the first approach is to make each GPU thread responsible for all dissimilarity calculations of a single region towards all its spatially adjacent regions or all non-spatially adjacent regions in the image. The second approach is to make each GPU thread responsible for the calculation of dissimilarity between only two regions, either spatially or non-spatially adjacent. The first approach takes a sequential behavior for the calculation of all dissimilarities for a specific region to its adjacent and non-adjacent regions, while other regions calculations are done in parallel. Thus the first approach doesn't take full advantage of parallel GPU threads. However, the second approach results in a much broader parallelism because it allows all dissimilarities of any regions pairs to be computed in parallel at the same time, making use of the complete independence of region-pair measurements, and no sequential calculation is needed. Figure 4.5 and Figure 4.6 show the difference between the two approaches.

For GPU implementations, many development platforms were considered such as; OpenCL [46] , NVidia Compute Unified Device Architecture (CUDA) [15] and Microsoft C++ Accelerated Massive Parallelism (C++ AMP) [16]. The main factors that were given the highest priority for selecting the desired platforms were platform maturity and flexibility. Thus two of these platforms were selected; NVidia CUDA and Microsoft C++ AMP, those were the two most mature and advanced GPU platforms that also provide top computation performance. GPU RHSEG is implemented using these two platforms for both approaches 1 and 2.

The RHSEG dissimilarity calculation is carried out in two stages in each iteration step. The first stage is the dissimilarity between every region and their spatially adjacent regions; this stage is called the spatial stage. Then the second stage is the dissimilarity measure between every region and all other non-adjacent regions, which is called spectral stage. Figure 4.2 shows both stages in HSEG flow chart. In both GPU approaches, each stage has a separate kernel, the spatial kernel and the spectral kernel. The spectral stage is the most computationally demanding task contributing to more than 95 % of total running time. To give a comprehensive overview of the GPU kernel implementation details, Figure 4.7 illustrates in detail how the GPU Approach 2 spectral kernel works and how regions are represented in the GPU memory.

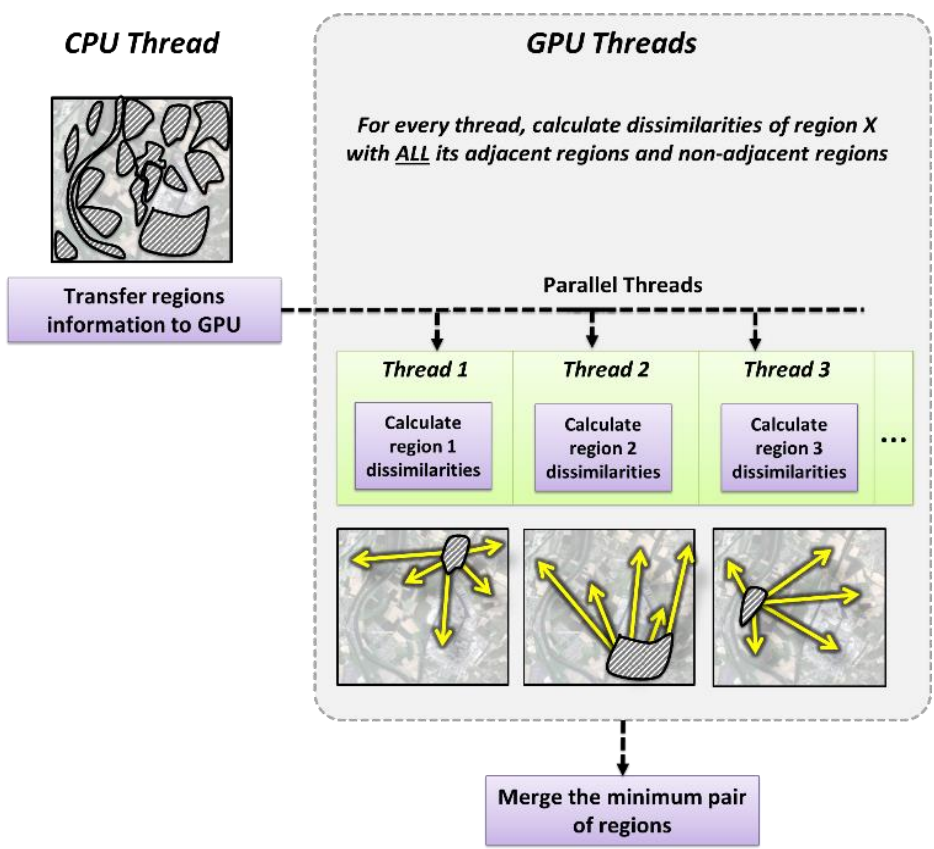

Figure 4.5. GPU Approach 1 (first GPU parallelization approach). Each GPU thread is responsible for the calculation of all dissimilarities for a certain region.

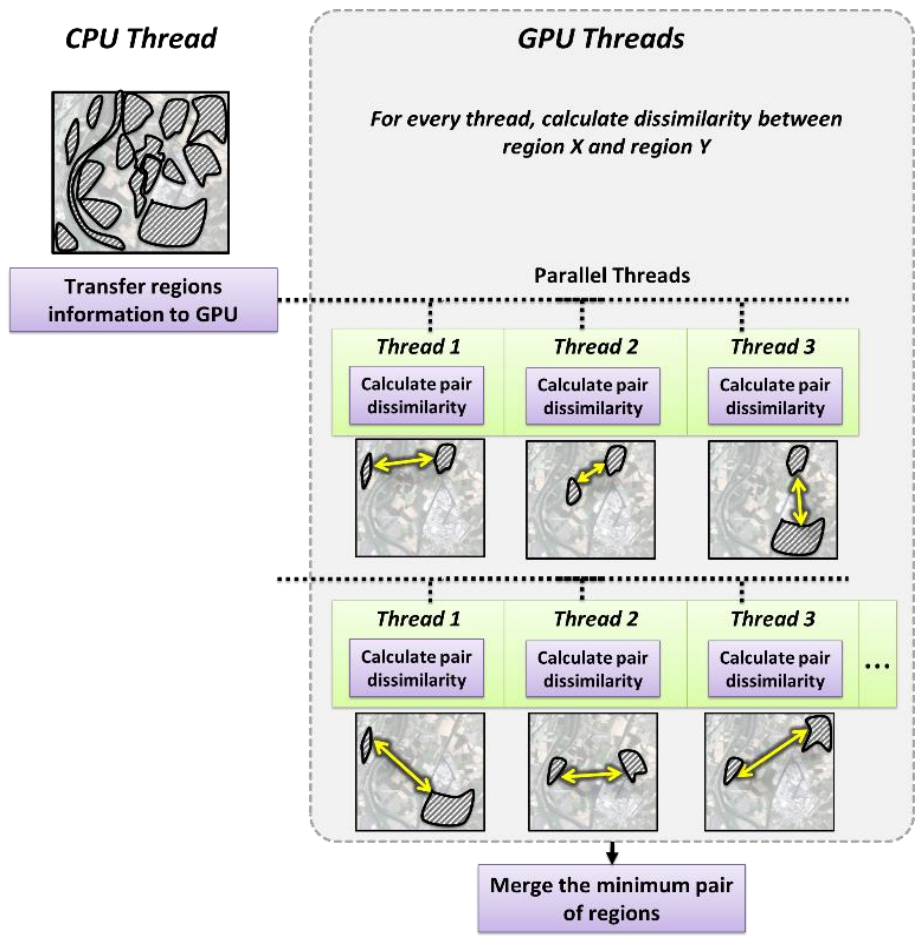

Figure 4.6. GPU Approach 2 (second GPU parallelization approach). Each GPU thread is responsible for the calculation of dissimilarity of only one pair of regions

In Figure 4.7, a sample image of size 6 x 6 pixels is passed to the spectral kernel. Before the kernel starts working, every pixel is considered a separate region, which gives 6 x 6 = 36 regions (this is only done once at the start of RHSEG, the next iteration uses the produced regions instead of image pixels). Then every region gets a unique ID from 1 to 36 and all the regions information (adjacent regions, spectral values of bands and number of pixels) is transferred to the GPU. The spectral kernel uses three arrays. First, the "Adjacencies" 2-dimensional array that is (number of regions) x (max_adjacencies) matrix is of type integer. It stores adjacent regions IDs of all regions, and allows each region to know its adjacent regions by their regional ID. Second, the "Pixels_Count" that is an array of the number of regions of type integer. It stores the number of pixels for every region. Finally, the "Bands_Sums" that is a matrix of (number of regions) x (bands) and stores the sum of region's pixel values at every band for all regions. The first two arrays reside in the GPU global memory, and the last one resides completely in the global memory and partially in the shared memory (for faster memory access). Finally, a fourth array is needed and is called "Best_Dissim". It stores the best dissimilarity value found for every region against all other regions.

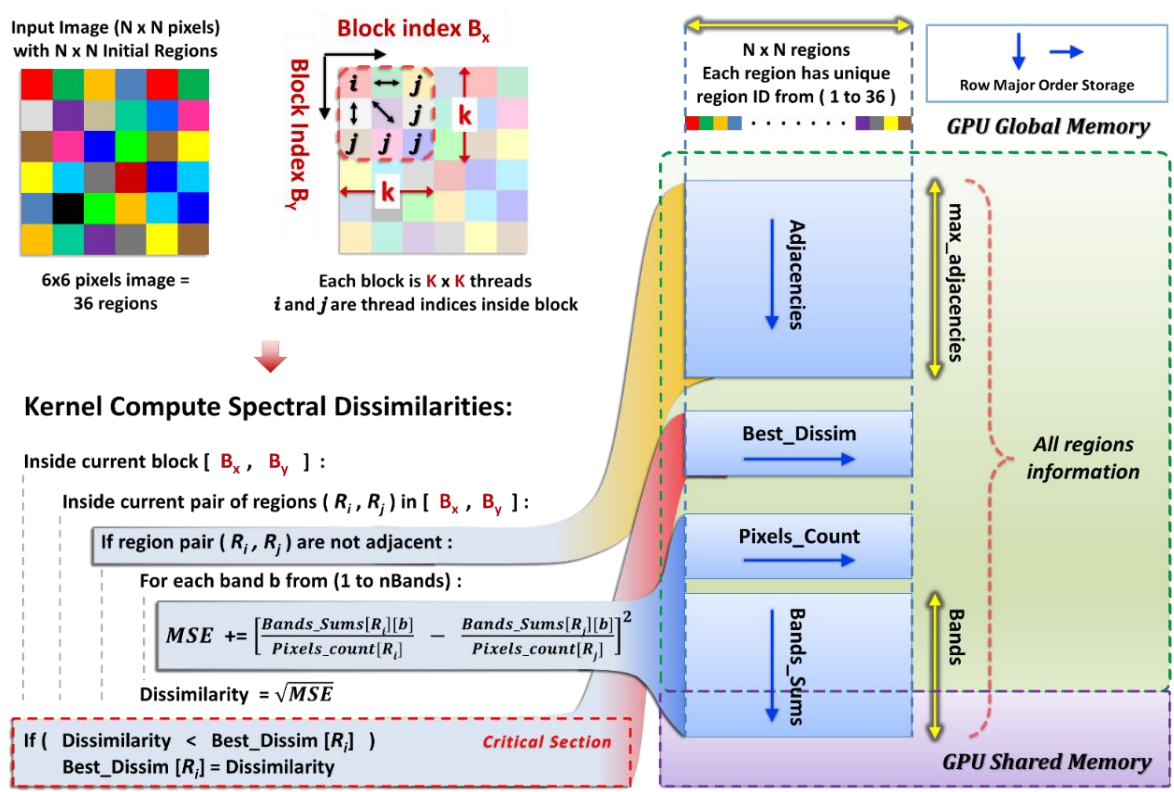

Figure 4.7. Example of spectral stage dissimilarities calculation for Approach 2 using GPU. The spectral kernel operates on N x N image using blocks of size K x K. GPU arrays that hold the required information for all regions. Dissimilarity equals square root of Band Sum Mean Square Error (MSE).

For optimizing memory access bandwidth, GPU on-chip shared memory is used. A small part of the "Bands_Sums" array is stored in every block's shared memory and the rest are accessed from the global memory. With the increase of GPU numbers of streaming multi-processors and shared memory size, more speedup can be achieved by using more shared memory. Figure 4.8 shows the detailed GPU code for approach 2 spectral kernel illustrated in Figure 4.7, which is called "kernel_compute_spectral_dissims". The details about shared memory size, kernel registers size and achieved threads occupancy are reported in chapter 5.

```
__global__ void kernel_compute_spectral_dissims ( const int* Adjacencies, const float* Bands_Sums
    , const int* Pixels_Count, float* Best_Dissim, int* Best_Dissim_Labels, int nRegions, const int nBands)
{
    int R_i_index = blockIdx.y * blockDim.y + threadIdx.y;
    int R_j_index = blockIdx.x * blockDim.x + threadIdx.x;

    __shared__ float shared_R_i_sums[BLOCK_WIDTH][BLOCK_BANDS];
    __shared__ float shared_R_j_sums[BLOCK_WIDTH][BLOCK_BANDS];

    //Load BLOCK_BANDS band sums from global "Bands_Sums" matrix
    // to shared memory
    For ( int b = 0; b < BLOCK_BANDS; ++b ) {
        shared_R_i_sums[threadIdx.y][b] =
                        Bands_Sums[R_i_index*nBands+b];
    }
    //...
    For ( int b = 0; b < BLOCK_BANDS; ++b ) {
        shared_R_j_sums[threadIdx.x][b] =
                        Bands_Sums[R_j_index*nBands+b];
    }
    __syncthreads();

    //Check if R_i and R_j are not adjacent
    int foundAdjacent = 0;
    for(int a = 0; a < MAX_ADJACENCIES; ++a) {
        if(Adjacencies[R_i_index*MAX_ADJACENCIES+a]
                                    == R_j_index) {
            foundAdjacent = 1;
            break;
        }
    }
    __syncthreads();

    float MSE = 0.0f;    float Dissimilary = MAX_FLOAT;

    if(foundAdjacent == 0) {
        float R_i_nPixels = Pixels_Count[R_i_index];

        float R_j_nPixels = Pixels_Count[R_j_index];

        //Calculate MSE using Regions means, first using those in
        //shared memory
        int band = 0;
        for ( band = 0; band < BLOCK_BANDS; ++band ) {
            float temp = shared_R_i_sums[threadIdx.y][band] /
                                                    R_i_nPixels
                       - shared_R_j_sums[threadIdx.x][band] /
                                                    R_j_nPixels;

            MSE += temp * temp
        }

        //and finally, using those in global "Bands_Sums"
        for ( band = band; band < nBands; ++band ) {
            float temp = Bands_Sums[R_i_index*nBands+band] /
                                                    R_i_nPixels
                       - Bands_Sums[R_j_index*nBands+band] /
                                                    R_j_nPixels;

            MSE += temp * temp;
        }

        Dissimilary = sqrtf(MSE * (R_i_nPixels*R_j_nPixels) /
                                        (R_i_nPixels+R_j_nPixels) );
    }
    // Critical Section =============================
    If ( Dissimilary < Best_Dissim[R_i_index] ) {
        Best_Dissim[R_i_index] = Dissimilary;
        Best_Dissim_Labels[R_i_index] = R_j_index;
    }
    // End of Critical Section ======================
}
```

Figure 4.8. RHSEG GPU Approach 2 spectral kernel.

In GPU, each block is composed of group of threads. The spectral kernel starts traversing all N x N regions using blocks of K x K threads in parallel, therefore the total number of bocks = N/K x N/K. In each block, dissimilarity between all regions inside the block is calculated as shown in Figure 4.7. The spectral kernel checks for every region pair ($R_i$, $R_j$) if they are not-adjacent, if true, it calculates the band sum mean square error value (BSMSE) over all the bands of the two regions, then the final dissimilarity is the square root of BSMSE. After the calculation of dissimilarity, the kernel needs to update the "Best_Dissim" array if it finds that the calculated dissimilarity is the smallest one so far for region $R_i$. Updating the "Best_dissim" array needs to be done "atomically" using a spin lock critical section to be carried out correctly. After the kernel is finished with all dissimilarity calculations for all regions in the input image, a GPU reduction step over "Best_Dissim" array is executed to find a pair or regions with minimum dissimilarity to be merged into one region. The two kernels (spatial then spectral) are then launched again after the merge is done to find new region pairs to merge. The process continues until the number of regions reaches the desired number of classes for the input image.

Several optimization techniques are taken into account in the design of either the sequential CPU or parallel GPU implementations. All proposed implementations are memory access optimized to improve the data locality and the cache memory hits. For Example, all arrays are accessed in row-major order, which is the sequential order of the byte arrangement in the CPU and GPU memory, and all arrays that reside in the GPU global memory have coalescent memory access. In addition, all the proposed implementations are accessed in blocks of K x K elements to improve the data locality. The proposed parallelized parts of RHSEG, that are calculating dissimilarities for each step and choosing the minimum pair to merge, contribute to more than 95% of the total execution time for both sequential and parallel of RHSEG. Other parts of the algorithm represent less than 5% of the execution time and are not suitable for parallel implementation, like merging a pair of regions after each step and stitching image sections for every recursive level.

In the following two sections, the proposed solutions for executing RHSEG algorithm using both multicore-CPU and GPU cooperatively is presented and is called Hybrid RHSEG. Section 4.3 describes implementation of RHSEG using a single multi-core CPU and a single GPU, while Section 4.4 describes the Hybrid RHSEG execution on a multi-node computer cluster of multi-core CPUs and GPUs.

For a multi-node hybrid cluster, RHSEG algorithm, an Amazon Elastic Compute Cloud (EC2) service is used [18]. However, C++ AMP is not currently capable of running on an Amazon network cluster because EC2 compute instances does not support running DirectX. Thus, the implementation of both parallel approaches of RHSEG algorithm on network clusters is implemented using a CUDA platform.

## 4.3    Hybrid CPU/GPU Parallel Implementation Using Single Computing Node

In RHSEG, for each recursive level the image is partitioned into 4 sections and this partitioning is repeated again for each quarter recursively till the deepest recursive level is reached. This means that for a 3 level RHSEG, there will be $4^2$= 16 image sections. The hybrid CPU/GPU implementation of RHSEG is based on distributing different image sections being processed at any level to the GPU and CPU cores. Therefore, different image section computations are executed in parallel on either a GPU or a CPU core. Besides, the algorithm is designed to work cooperatively; a CPU core can pass its image section to a GPU if it is free, thus GPU can help in finishing the computation faster and achieve the best utilization. Figure 4.9 shows the parallel execution of RHSEG on hybrid CPU/GPU with a 4 core CPU and single GPU.

In Figure 4.9, the execution starts with the deepest level of recursion, where we have 4 indivisible image sections ready for HSEG computation. The four image sections are distributed to GPU and CPU as follows: Image section one goes to the GPU and one CPU core (thread); sections 2, 3 and 4 go to the other 3 CPU cores (as threads). In this way, the computation of the 4 sections is executed in parallel. The GPU thread is already faster than any CPU thread; therefore, when the GPU has finished its image section, it is considered free to conduct future computations of any other image sections. This allows RHSEG to assign a computation of any other image sections to the GPU. Therefore, the GPU picks up any remaining image section that has not been processed. If all image sections are being processed, it picks up an image section from any running CPU thread to finish it faster. On the other hand, if the GPU finishes the current image section, then it repeats the same technique by finding another image section to compute, until all images sections are finished. A control thread always looks for every 4 image sections finished of certain level, then it combines their results, and the algorithm terminates when the control thread combines the results for the first level (level 1). Algorithm 1 illustrates a Hybrid CPU/GPU RHSEG implementation.

---

**Algorithm 1: Hybrid CPU/GPU RHSEG**

---

*Input*: ($f$, $L$) where $f$ is image with N x N pixels, $L$ is number of desired RHSEG recursion levels
  1: Intialize $GPU\_Ready$ = true, $nSections = 4^{(L-1)}$ , array $Migrated\_to\_GPU$ [$nSections$] = [false]

  2: **procedure** Hybrid_RHSEG_Thread ($S_i$)        {$S_i$ is an image section of input image $f$}
  3:     $Migrated\_to\_GPU$ [$S_i$] = false
  4:     **while** desired number of clusters not reached **do**
  5:         **if** $Migrated\_to\_GPU$ [$S_i$] = true **then**
  6:             $GPU\_Ready$ = false
  7:             **GPU_HSEG** ($S_i$)
  8:         **else**
  9:             **CPU_HSEG** ($S_i$)
 10:         **end if**
 11:     **end while**
 12:     **if** $Migrated\_to\_GPU$ [$S_i$] = true **then** $GPU\_Ready$ = true
 13: **end procedure**

 14: **procedure** Control_Thread ($f$, $L$)
 15:     partition $f$ to equal $4^{(L-1)}$ image sections [ $S_1 - S_{4^{(L-1)}}$ ]
 16:     $Q = [S_1 - S_{4^{(L-1)}}$ ]            { put all image sections in queue $Q$ for processing }
 17:     **while** $Q$ not empty **do**
 18:         $S_i$ = pop image section from $Q$
 19:         Hybrid_RHSEG_Thread($S_i$) {Create new hybrid thread and send $S_i$ as input}
 20:     **end while**
 21:     **while** not all threads are finished **do**
 22:         **if** $GPU\_Ready$ = true **then**
 23:             **for every** thread $t$  in all running threads **do**
 24:                 **if** $S_i$ used in $t$ is processed on CPU core **then**
 25:                     $Migrated\_to\_GPU$ [$S_i$] = true
 26:                     $GPU\_Ready$ = false
 27:                 **end if**
 28:             **end for**
 29:         **end if**
 30:     **end while**
 31: **end procedure**

---

To guarantee the scalability by increasing the number of CPU cores, the algorithm is designed to dynamically use any free available cores for requested image section computations. For example, if an 8 core CPU is used, then each core of the 8 cores receives an image section from the control thread. Then the computation is carried out for each section. After that, the results return to the control thread and the 8 cores will be free to process any other image sections. The control thread is responsible for dispatching image sections to threads and receiving results from different threads for combining at different levels. Figure 4.10 illustrates the execution process of RHSEG on 8 CPU cores and one GPU.

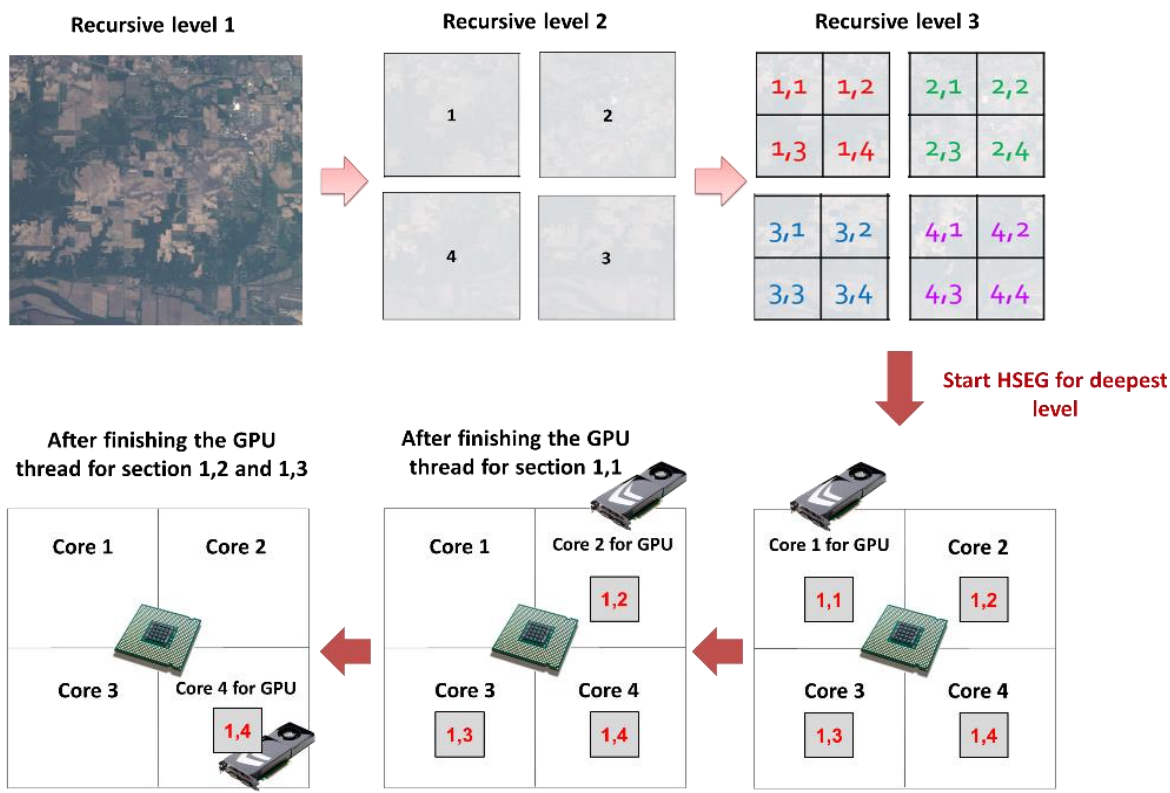

Figure 4.9. Step by step Hybrid CPU/GPU RHSEG with 3 recursive levels using 4 cores CPU, computation starts at the deepest third level.

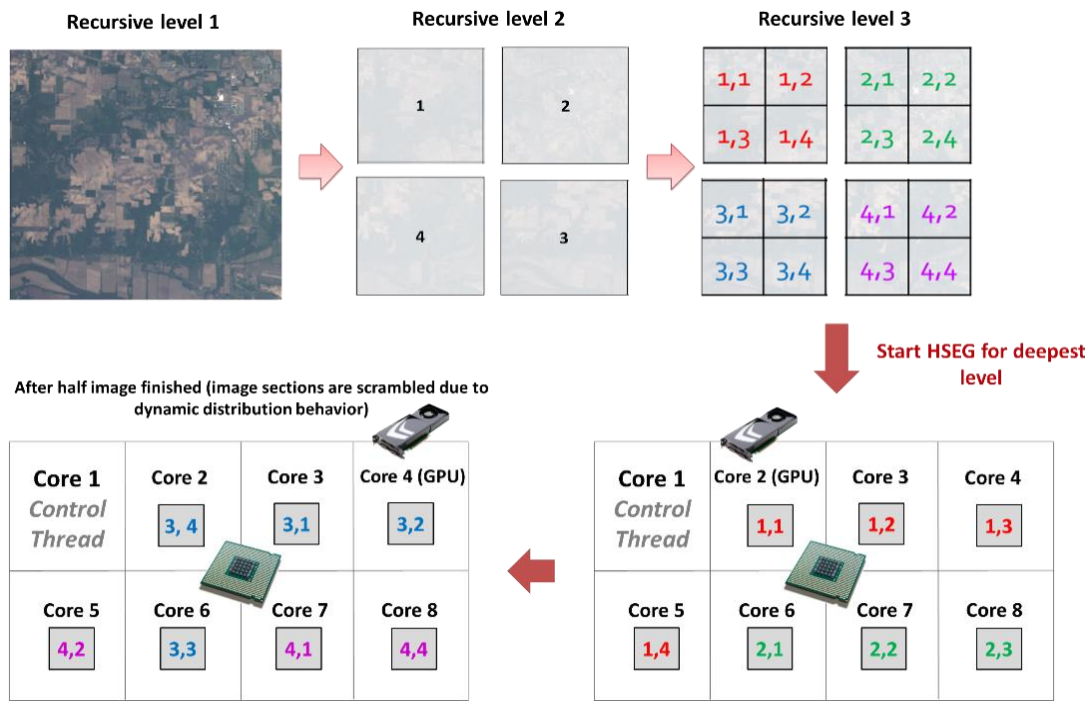

Figure 4.10. Hybrid RHSEG using 8 CPU cores and one GPU.

## 4.4 Hybrid CPU/GPU Parallel and Distributed Implementation on Multi-Node Computer Cluster

This section describes multi-node cluster distributed implementations of the RHSEG algorithm; hybrid CPU/GPU cluster, GPU cluster, Multi-core CPU cluster and CPU Cluster. The distributed cluster technique of the hybrid RHSEG is similar to the technique described earlier for multi-core machines, but uses network nodes as the distributed computing element. Image sections are distributed to network nodes instead of CPU cores (threads) and the control thread of the master node receives section results and stitches them for any recursion level. The master node itself is also used as a computing node. For example, in Figure 4.11 four cluster nodes are used. Each node in the cluster has 8 CPU cores and one GPU. The 8 CPU cores in each node are used for the computation of the dedicated image sections sent to this node.

The GPU cluster implementation of RHSEG is similar to the technique described earlier for hybrid CPU/GPU clusters but without cooperation of multi-core CPUs. The control thread allows the GPU only to process the images sections from the queue. Therefore at each node, the GPU alone is working, and no CPU core is used for computation. Similarly, the Multi-core CPU Cluster implementation of RHSEG works just as the hybrid CPU/GPU cluster technique, but the GPUs are not allowed to work or process any image sections. Finally, the CPU cluster implementation works just as the hybrid cluster but with only a single CPU core in each network node allowed to process image sections.

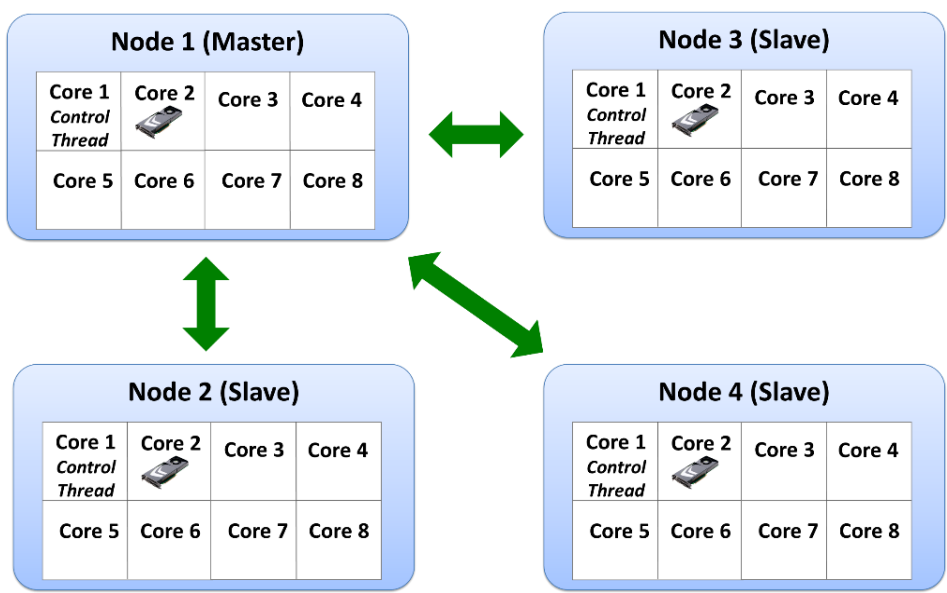

Figure 4.11. Example of cluster Hybrid RHSEG, 4 cluster nodes (each one consists of 8 CPU cores and single GPU).

# Chapter 5

# Experiments and Results

# 5    Experiments and Results

## 5.1    Evaluation Methodology

Three categories of experiments were carried out to study the different proposed parallelized versions of the RHSEG algorithm; accuracy assessment, execution time and energy consumption experiments. First, accuracy assessment shows the classification accuracy of selected data set against the ground truth information for both parallel and sequential CPU and GPU solutions. Second, sets of execution time experiments are conducted to study the speedup of the proposed parallel CPU/GPU solutions compared to sequential CPU solution. The experiments were carried out under different parameters and data configurations that affect the execution time, such as; image size, image depth (number of bands), image details and number and the dimensions of GPU threads. Finally, energy consumption experiments show the power/energy consumption rates of GPU solutions compared to both sequential and parallel CPU solutions.

For execution time experiments, three sets of experiments were conducted. First, parallel RHSEG on a single GPU without a multi-core CPU is carried out using both CUDA and C++ AMP technologies. Second, parallel RHSEG using a Hybrid CPU/GPU single computing node is carried out. Finally, parallel RHSEG using different multi-node clusters are carried out; GPU cluster, hybrid CPU/GPU cluster, CPU cluster and Multi-core CPU cluster. The performance of the parallel implementation is measured by calculating the speedup, which is the number of times a parallel implementation is faster than the sequential one on a single CPU core.

### 5.1.1 The Data Set

The experiments are performed using five different images, three real hyperspectral images and two manually synthetic images. The three hyperspectral images are; the Indian Pines AVIRIS hyperspectral data [58], the Pavia Center and Pavia University data. Figure 5.1 shows portions of these hyperspectral images. The Indian Pines scene was gathered by the AVIRIS instrument. It consists of 16 ground truth classes. It was acquired over a mixed agricultural/forested region in NW Indiana. Four noisy bands were removed and the rest of 220 spectral bands are used. Pavia data in Italy was collected by the ROSIS [59] sensor. The first image was collected over Pavia city center, Italy. It contains 102 spectral channels and 9 ground truth classes. The second image was collected over the University of Pavia with nine ground-truth classes and 103 spectral bands.

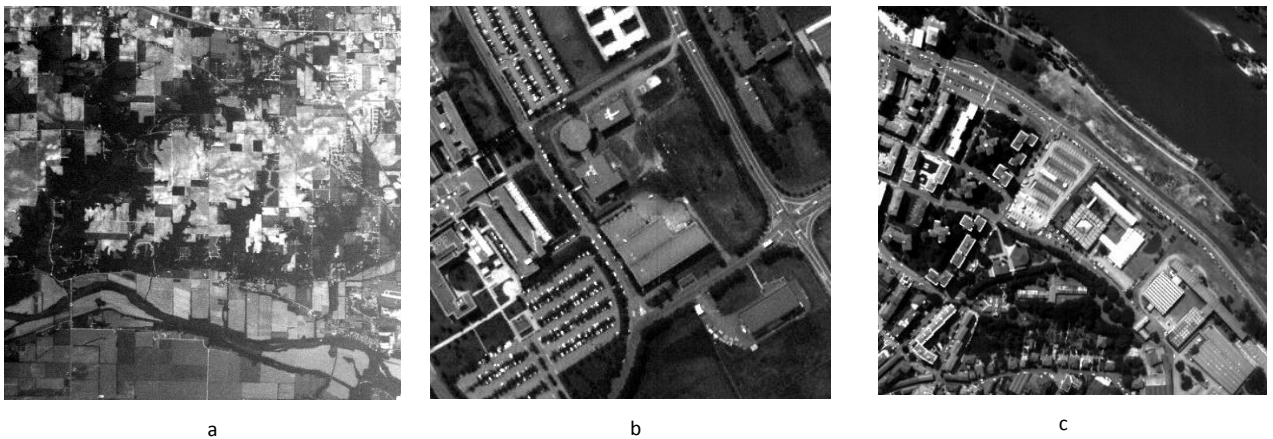

a                                b                                c

Figure 5.1. a)  Indian Pines Data Set, b) Pavia Center Data Set, c) Pavia University Data Set.

Experiments are carried out using different hyperspectral image sizes of 128x128, 256x256 and 512x512 pixels. For each image size, data was cropped from the large image, not scaled. The number of bands for Indian

Pines image is 220 bands and for Pavia Center image and the University image is 102 and 103 respectively. The spectral clustering weight parameter used for all experiments is 0.21, this is an acceptable value that produces clear shaped classification results while not losing the recognition of the non-adjacent regions of image classes.

## 5.1.2  Hardware Architectures

The execution of non-hybrid single node (using only GPU without CPU cores) RHSEG algorithm is tested using an NVidia GeForce 550 Ti and Tesla M2050 devices. GeForce 550 Ti consists of 192 processing cores each operating at 1940 MHz, with 1024 MB GDDR5 192-bit memory interface, which operates at 2050 MHz that is capable of 98.4 GB/sec memory bandwidth. Tesla M2050 contains 448 processing cores each operating at 1147 MHz, with 384-bit memory operating at 1546 MHz clock. The CPUs used are Intel Core i5 with 3100 MHz, 256 K.B. L1, 1 M.B. L2 and 6 M.B. L3 cache memories (for GeForce 550 Ti) and Intel Xeon X5570 (for Tesla M2050). Table 5.1 summarizes hardware specification for non-hybrid experiments.

Table 5.1. Hardware Specifications for Non-Hybrid Sequential and parallel RHSEG Experiments.

|  | CPU 1 | GPU 1 | CPU 2 | GPU 2 |
|---|---|---|---|---|
| **Processor Name** | Intel Core i5 | NVIDIA GeForce 550 Ti | Intel Xeon X5570 | NVidia Tesla M2050 |
| **Number of Processors** | 1 CPU core | 192 processing cores | 1 CPU core | 448 processing cores |
| **Clock Speed** | 3100 MHz | 1940 MHz | 2.93GHz | 1147 MHz |
| **Memory Size/Bandwidth** | - 12 GB Memory (10.6 GB/sec)<br><br>- 256 K.B. L1, 1 M.B. L2 and 6 M.B. L3 cache | 1024 MB GDDR5 memory (98.4 GB/sec) | 22 GB | 2 GB (148 GB/ sec) |

For the multi-node hybrid cluster RHSEG algorithm, an Amazon Elastic Compute Cloud (EC2) [18] service is used. Each EC2 node used dual quad core (total 8 cores) Intel Xeon X5570 operating at 2.93 GHz, and the GPU is NVidia Tesla M2050 device running on Windows Server 2012. EC2 Cluster nodes are connected to each other by a 10 Gigabit/s Ethernet network. All implemented code was compiled using Microsoft Visual C++ 2012 with compiler flag /O2 for speed optimization. To ensure the consistency of square root floating point calculation across all different parallel and sequential architectures, appropriate compiler flags were used in all implementations to force accurate calculations based on the IEEE 754 floating point standard; For Visual C++ 2012 CPU sequential code, the flag /fp:precise was used,  for all CUDA parallel implementations the flags --prec-sqrt = true, --gpu-architecture = compute_20  and --gpu-code = sm_21 are used, finally for C++ AMP parallel implementation the precise math library namespace "Concurrency::precise_math" was used. Table 5.2 summarizes hardware specification for hybrid cluster experiments.

Table 5.2. Hardware Specifications of Amazon Elastic Compute Cloud (EC2) used for Multi-Node and Single-Node Hybrid Sequential and Parallel RHSEG Experiments.

|  | CPU | GPU |
| --- | --- | --- |
| **Processor Name** | Intel Xeon X5570 | NVidia Tesla M2050 |
| **Number of Processors** | 2 x 4 Core Processors | 448 processing cores |
| **Clock Speed** | 2.93GHz | 1147 MHz |
| **Memory Size** | 22 GB | 2 GB |

## 5.2 Experiments and Results

### 5.2.1 Accuracy Assessment

The accuracy assessment is needed to ensure that parallel solutions are correctly implemented and identically matches the sequential implementation. The classification accuracy assessment of the proposed parallelized versions of the RHSEG algorithm is carried out using a Pavia Center dataset. The image was cropped to size 490x490 pixels with 97 bands after removing first five noisy bands. Figure 5.2 shows a section of the Pavia center image and the corresponding ground truth image of nine classes. The classification result is compared to the provided ground truth information. Classification is carried out using GPU, hybrid CPU/GPU and sequential CPU solutions. In all three cases, the classification results were identical. Table 5.3 shows the RHSEG segmentation result using square root of BSMSE dissimilarity criterion with four levels of recursion and the spectral clustering weight equals 0.15. The coarsest segmentation result is selected that separates most of the nine classes. Each segmentation class was assigned to a specific ground truth class that covered the plurality of their pixels. Accuracy scores for all nine materials are shown in Figure 5.3 with an overall accuracy of 76%. Figure 5.4 shows both classification result and ground truth images with 16 classes of Indian Pines dataset.

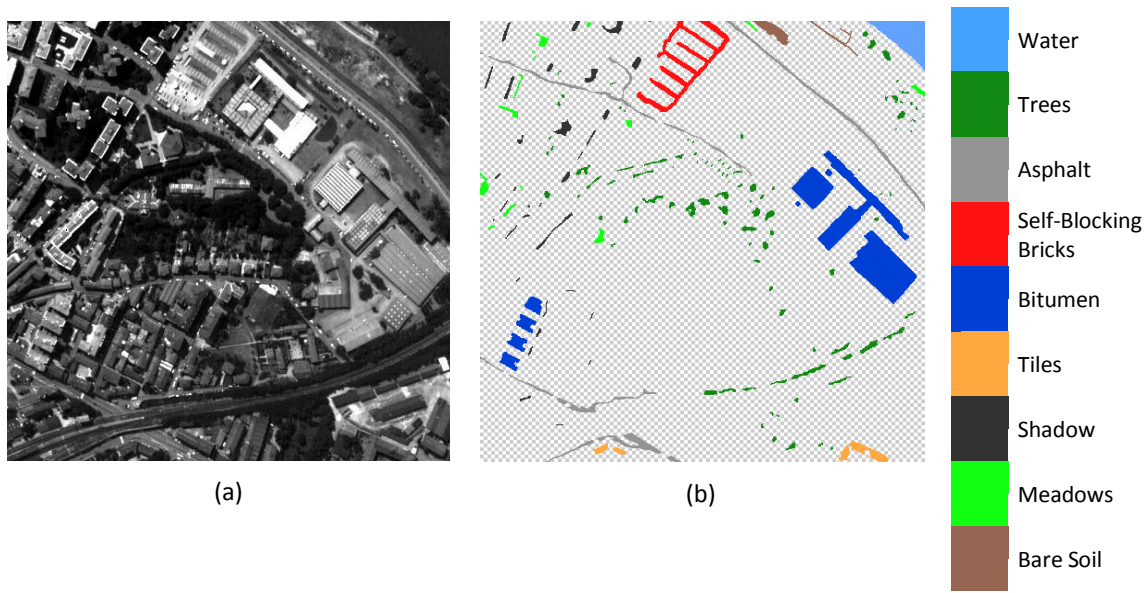

| | |
|---|---|
| | Water |
| | Trees |
| | Asphalt |
| | Self-Blocking Bricks |
| | Bitumen |
| | Tiles |
| | Shadow |
| | Meadows |
| | Bare Soil |

(a)       (b)

Figure 5.2. a) Pavia Center image section of 490x490 pixels containing all nine classes provided with the dataset, b) Pavia Center ground truth classes with color key for each class.

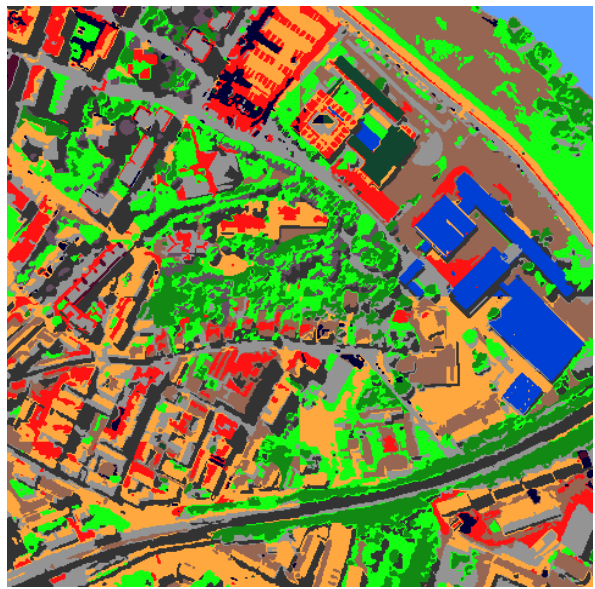

Figure 5.3 . Classification map for Pavia Center image section showing all nine ground truth classes

Table 5.3. Classification accuracy for each ground truth class of Pavia Center dataset

| Class | Accuracy % |
| --- | --- |
| Water | 100 |
| Trees | 62.7 |
| Asphalt | 59.9 |
| Self-Blocking Bricks | 68.2 |
| Bitumen | 84.3 |
| Tiles | 56.1 |
| Shadow | 99.7 |
| Meadows | 61.4 |
| Bare Soil | 92.3 |
| **Overall** | **76** |

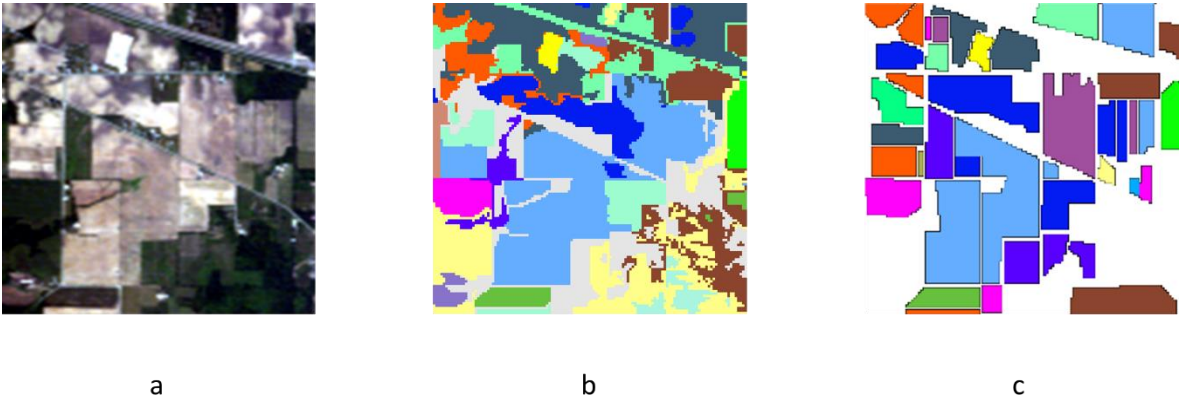

a                                  b                                  c

Figure 5.4.  a) Indian Pines Data Set RGB image of size 128x128 pixels, b) the classification map image
consists of 16 classes and c) the corresponding ground truth image with 16 classes

## 5.2.2   Non-Hybrid Single Node Experiments

### 5.2.2.1   GPU parallel RHSEG compared to Sequential RHSEG

This experiment is carried out to study the parallelized implementation of
RHSEG using a single a GPU. Figure 5.5 shows the execution times (in
seconds) of the RHSEG algorithm using different image sizes for both approaches implemented by CUDA and C++ AMP. For 128x128x220 image size, the RHSEG CPU sequential execution time is around 7920 seconds, while the CUDA GPU Approach 1 execution time is around 2486 seconds, C++ AMP approach 1 is 2180 seconds, CUDA approach 2 is 640 seconds and finally C++ AMP approach 2 is 930 seconds. One can see from Figure 5.5 that the proposed approaches to implement RHSEG using a single GPU have far less execution time than sequential implementations on a CPU.

The GPU running time includes the memory copy time between the main memory and the GPU memory. Table 5.4 shows the speedups of RHSEG parallel approach 1 and 2 on a single node GPU with respect to sequential implementation on a CPU using CUDA and C++ AMP platforms. A 3.1x and 3.5x average speedup is achieved for Approach 1 for CUDA and C++ AMP respectively and 12x, 8x and 21x average speedup for CUDA and C++ AMP Approach 2 over the sequential CPU implementation.

In this experiment, Approach 2 kernels were launched using a block size of 16 x 16 threads, the maximum size of on-chip shared memory used is 2KB per block. The spectral kernel in Figure 4.8 uses 24 registers in each thread, so that for 16x16 block a total of 6144 registers are used. The current implementation for Approach 2 using the described algorithm in CUDA results in 78% occupancy, which means that every streaming multiprocessor in the GPU runs 1200 threads at a time out of the maximum available 1536 per streaming multiprocessor (with GeForce 550 Ti device).

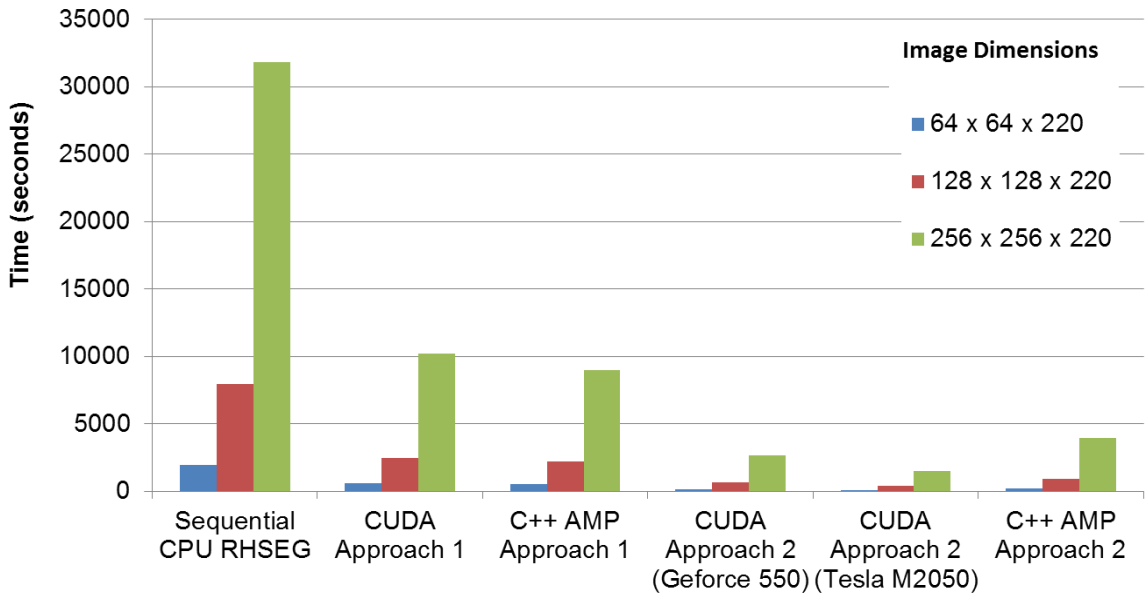

Figure 5.5. Execution times (in seconds) of RHSEG parallel Approach 1 and 2 using CUDA and C++ AMP on single GPU, for different image sizes.

Table 5.4. Speedups of RHSEG parallel approach 1 and 2 on Single node GPU with respect to sequential implementation on CPU.

| Image Dimensions Width x Height x Bands | CUDA GPU Approach 1 Speedup | C++ AMP Approach 1 Speedup | CUDA GPU Approach 2 Speedup (GeForce 550 Ti) | CUDA GPU Approach 2 Speedup (Tesla M2050) | C++ AMP Approach 2 Speedup |
|---|---|---|---|---|---|
| 64x64x220 | 3.2x | 3.7x | 12.6x | 21.8x | 8.9x |
| 128x128x220 | 3.1x | 3.6x | 12.3x | 21.7x | 8.5x |
| 256x256x220 | 3.1x | 3.5x | 12.1x | 21.6x | 8.0x |
| 512x512x220 | 3.0x | 3.4x | 11.8x | 21.5x | 7.6x |

### 5.2.2.2   Impact of Image Details on Speedup

This experiment was performed to study the impact of changing the image details on the achieved speedups of RHSEG using a single GPU. Figure 5.6 shows three images that differ in details. Figure 5.6(a) and Figure 5.6(b) are synthetic images generated manually for the sake of the experiment. Figure 5.6(c) is a portion of the Indian Pines image. The images are differing in the number of classes/regions. Each image size is 50x50 pixels x 220 bands. Table 5.5 shows the speedup of RHSEG on single GPU using different images with different details. One can see from Table 5.5 that the speedup almost is not affected by increasing the number of region/classes. Then changing the complexity and details of the image does not affect the speedup significantly.

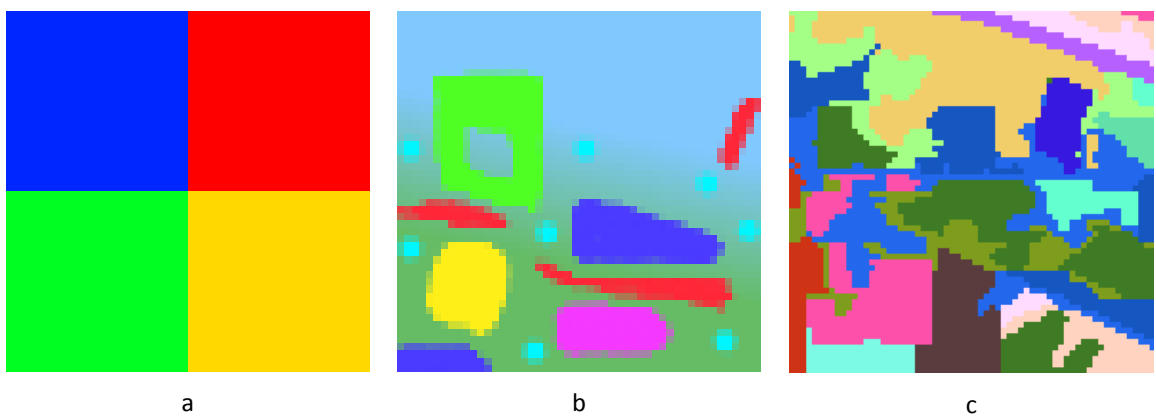

a                                   b                                   c

Figure 5.6. a) Detail Image 1: Synthetic image with 4 classes/4 regions, b) Detail Image 2: Synthetic image with 8 classes/12 regions, c) Detail Image 3: Portion Indian Pines image with 16 classes/25 regions.

Table 5.5. Speedups of RHSEG on single GPU (CUDA and AMP for Approaches 1 and 2 respectively) using different image details with respect to sequential implementation on CPU

| Image Details | Single GPU Speedup (CUDA / C++ AMP Approach 1) | Single GPU Speedup (CUDA /C++ AMP Approach 2) |
|---|---|---|
| Image a (4 classes / 4 regions) | 3.1x / 3.8x | 12.7x / 9.5x |
| Image b (8 classes / 12 regions) | 3.1x / 3.8x | 12.7x /9.5x |
| Image c (16 classes / 25 regions) | 3.3x / 3.9x | 12.8x / 9.6x |

### 5.2.2.3  Impact of Image Depth on Speedup

This experiment was performed to study the impact of changing the image depth (number of bands) on the execution time of RHSEG using a single GPU.

For an image size of 32x32 pixels, the experiments are carried out using 3, 10, 50, 100, 150 and 220 bands. Table 5.6 shows the performance of the GPU implementation for different numbers of bands. For GPU Approach 1, the speedup increases slightly by increasing the number of bands. On the other hand, with GPU Approach 2, the speedup increases significantly by increasing the number of bands. GPU Approach 2 with three bands achieves 2x speedup while using 220 bands achieves 12x speedup with respect to sequential CPU. Hence, it is clear that both the CPU approaches are significantly sensitive to changing number of bands.

Table 5.6. Speedups of RHSEG on single GPU (CUDA and C++ AMP for Approaches 1 and 2 respectively) using different image depths with respect to the sequential implementation on CPU

| Image Depth (# of Bands) | Single GPU Speedup (CUDA/ C++ AMP) Approach 1 | | Single GPU Speedup (CUDA/ C++ AMP) Approach 2 | |
|---|---|---|---|---|
| 3 | 1.3x | 0.1x | 2x | 0.09x |
| 10 | 2.8x | 0.4x | 6.5x | 0.3x |
| 50 | 3.0x | 2.2x | 11.4x | 1.5x |
| 100 | 3.3x | 3.0x | 12.5x | 2.8x |
| 150 | 3.3x | 3.5x | 13x | 7.3x |
| 220 | 3.3x | 3.9x | 12.8x | 9.6x |

### 5.2.2.4  Impact of GPU Thread Block Size on Speedup

This experiment was performed to study the effect of changing the number of threads per block for a single GPU for both GPU Approaches 1 and 2. For an image size of 32x32 pixels x 220 bands, the experiments are carried out using 4x4, 8x8, 16x16 threads per block for approach 2 (CUDA and C++ AMP). Table 5.7 shows the performance of the GPU implementation for the different number of threads per block.  It is noticeable that changing the block size affects the speedups. For example, speedups increased significantly by increasing block size from 4 x 4 to 16 x 16. The optimal block size for given inputs was 16 x 16 threads per block.

Table 5.7. Speedups of RHSEG on single GPU (CUDA and C++ AMP for Approaches 1 and 2 respectivley) using different thread per blocksizes with respect to sequential implementation on CPU

| GPU Threads per block | Single GPU Speedup (CUDA / C++ AMP) |
|---|---|
| 4x4 threads | *N/A* / 5.3x |
| 8x8 threads | 8.4x / 8.9x |
| 16x16 threads | 12.8 / 9.6x |

## 5.2.3  Hybrid Single Node CPU/GPU RHSEG

This experiment was performed to measure speedups of parallelized implementation of RHSEG Approach 2 on a single node hybrid CPU/GPU using CUDA. For 64x64x220 image size, the RHSEG CPU sequential execution time is around 2033 seconds, while the RHSEG GPU execution time is around 94 seconds and the hybrid parallel execution time was about 89 seconds. Table 5.8 shows speedup results for a GPU node and single hybrid CPU/GPU node against sequential implementation on a CPU. A 21.6 and 22.8x average speedup is achieved for a single GPU and hybrid CPU/GPU implementation respectively over the sequential CPU implementation.

Table 5.8. Speedups of RHSEG algorithm on Single node using GPU or Hybrid CPU/GPU with respect to sequential implementation on CPU

| Image Dimensions | GPU | Hybrid CPU/GPU (8 CPU Cores) |
|---|---|---|
| 64x64 | 21.8x | 22.8x |
| 128x128 | 21.7x | 22.9x |
| 256x256 | 21.6x | 22.8x |
| 512x512 | 21.5x | 22.7x |

## 5.2.4  Hybrid Multi-Node Cluster CPU/GPU RHSEG

This experiment was performed to measure speedups of parallelized RHSEG on different multi-node cluster types, GPU cluster, hybrid CPU/GPU multi-node cluster, CPU cluster and multi-core CPU cluster. Execution times are recorded and compared with the CPU sequential execution time. Also the execution time is compared with the single GPU implementation. In this experiment, NVidia Tesla M2050 is used for both single and multi-node GPU clusters and hybrid clusters. For the Indian Pines image, the experiments are carried out using 256x256x220 and 512x512x220 pixels x bands. Table 5.9 shows the results for 4, 8 and 16 cluster nodes. Figure 5.7 shows the speedup expressed as a function of the number of nodes for the Indian Pines image of size 512x512 pixels. One can observe from Figure 5.7 that the speedup increases by increasing the number of nodes. Furthermore, one can observe from Table 5.9 that a speedup of 15, 55, 249 and 259 times on a CPU cluster, multi-core CPU cluster, GPU cluster and hybrid CPU/GPU multi-node cluster respectively are achieved over the sequential CPU implementation.

Table 5.9. Speedups of RHSEG on multi node Hybrid CPU/GPU Cluster with respect to sequential implementation on CPU, CPU Cluster, and Multicore CPU cluster

| Image Size | No. of nodes | CPU Cluster | Multicore CPU cluster (8 Cores) | GPU Cluster (NVidia Tesla M2050) | Hybrid CPU/ GPU Cluster | Single GPU (NVidia Tesla M2050) |
|---|---|---|---|---|---|---|
| 256x256 | 4 | 3.9x | 29x | 80x | 84x | 21.6x |
| | 8 | 7.8x | 55x | 146x | 153x | |
| | 16 | 15.4x | 55x | 249x | 259x | |
| 512x512 | 4 | 3.9x | 30x | 78x | 82x | 21.5x |
| | 8 | 7.7x | 57x | 140x | 146x | |
| | 16 | 15.1x | 106x | 232x | 241x | |

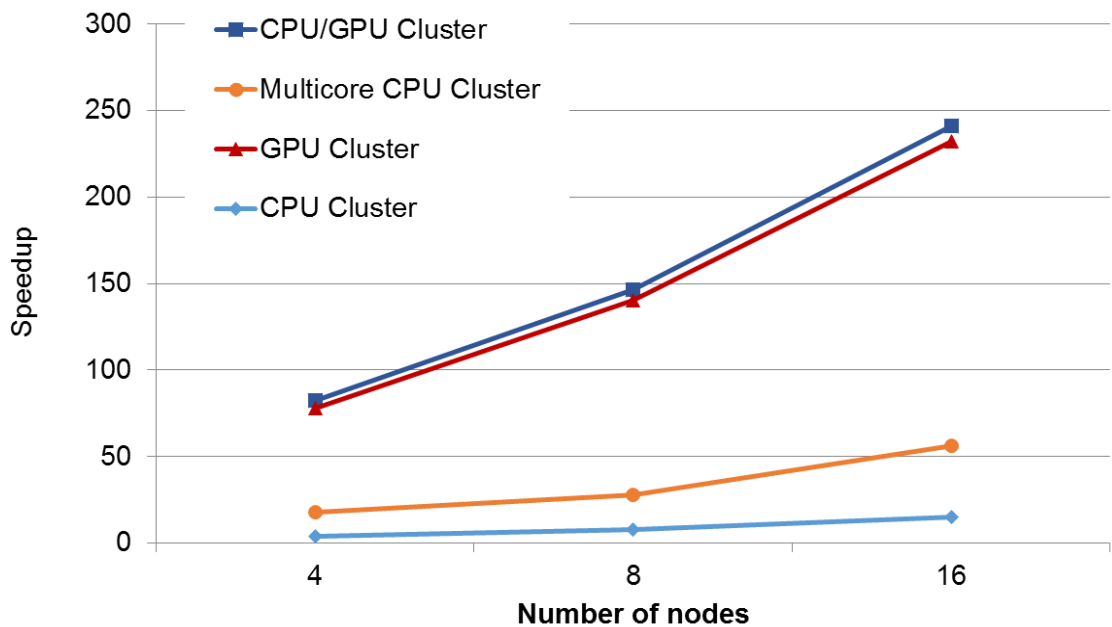

Figure 5.7. Hybrid CPU/GPU RHSEG cluster speedups of different cluster sizes: 4, 8 and 16 nodes

## 5.2.5  Power Consumption

Finally, the last experiment was performed to study the power/energy consumption of the proposed parallel GPU/CPU solutions against the sequential CPU solution. A power meter device is used to read the Watts consumed by the CPU unit from the wall socket, thus the samples from the power meter are collected externally and separately from the experiment system, in order to prevent the measurements from affecting the accuracy of the experiments results. Power and energy consumed by the system in an idle state (i.e. disks, fans and idle CPU/GPU processing) is measured separately and subtracted from the computation measurement results. During experiment execution, power readings decrease over time, so the power readings from the meter are collected over execution duration and the average power and energy are calculated and used for the comparison. Figure 5.8 shows the KD302 [60] [61] power meter device used for these experiments.

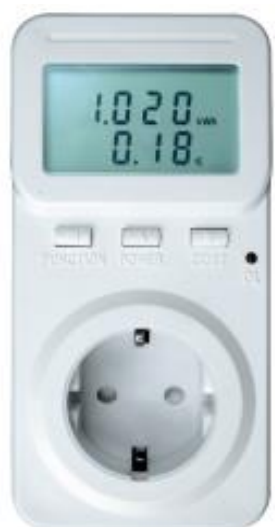

Figure 5.8. The KD302 power meter device used for power measurements

The power and total energy consumed are measured during the computing period of both CUDA and C++ AMP approach 2 computations on 128x128x220 image size. The average power and energy consumption values are calculated by taking the mathematical average of five repetitive power and energy measurements for every experiment. Table 5.10 shows the power and energy consumption measurements for both CUDA and C++ AMP on a single NVidia GeForce 550 Ti GPU, column four and six show the relative energy consumption ratio of the different parallel GPU platforms to both serial and parallel CPU.

Table 5.10. Single GPU energy consumption for CUDA and C++ AMP Approach 2 compared to sequential and parallel CPU energy consumption

| | Image Size (Width x Height x Bands) | Average power consumption (Watts) | Average energy consumption (Joules) [Power x Time] | GPU energy consumption compared to sequential CPU % | Equivalent parallel CPU energy consumption (Joules) [same GPU speedup] | GPU Energy consumption compared to equivalent parallel CPU cluster % |
|---|---|---|---|---|---|---|
| CPU Sequential RHSEG | 128x128x220 | 15 | 117,600 | N/A | N/A | N/A |
| Approach 2 (CUDA) | 128x128x220 | 115 | 69,920 | 59% | 80,260 | 88% |
| Approach 2 (C++ AMP) | 128x128x220 | 75 | 60,000 | 52% | 81,600 | 74% |

From Table 5.10, it is noticeable that Approach 2 clearly achieves less energy consumption than the sequential CPU solution. It is more useful to compare the energy consumption of the proposed parallel GPU solutions against the parallel CPU solution, not only the sequential one, so that we can decide if it is beneficial in terms of energy consumption to use the GPU parallel system instead of the parallel CPU system. The last column in Table 5.10 shows the ratio of energy consumption of GPU platforms to equivalent parallel CPU cluster. The "Equivalent parallel CPU cluster" means that for a certain GPU platform speedup, a parallel CPU cluster is configured to achieve the same speedup, and then the power consumption of the two systems is compared. For example, for both CUDA and C++ AMP, a parallel CPU cluster of 4 and 3 computing nodes, each with four CPU cores achieving up to 12.8x and 9.6x speedups respectively is used, and their energy consumptions are calculated (excluding the idle power). Then the ratio of approach 2 (CUDA / C++ AMP) to the CPU cluster energy consumption is calculated. It is found that approach 2 CUDA and C++ AMP energy consumption is lower than the equivalent parallel CPU cluster by 12% and 26% respectively, a reduction from 100% to 88% and 74% respectively.

# Chapter 6

# Conclusions and Future Work

# 6    Conclusions and Future Work

## 6.1    Conclusions

This study proposed parallelized an RHSEG algorithm using graphical processing units (GPUs) with the co-operation of multi-core CPUs and computer clusters for onboard processing scenarios. RHSEG is a well-known object-based image analysis (OBIA) technique that is developed by NASA for effectively analyzing hyperspectral images with high spatial resolutions. The proposed parallel implementations are focused towards onboard processing by both accelerating execution time and reducing the power consumption by using GPUs that are lightweight computation devices with low power consumption potential for certain tasks.

Three parallel solutions are proposed; parallel RHSEG using a single GPU without a multicore CPU and implemented using both CUDA and C++ AMP technologies, parallel RHSEG using a Hybrid multicore CPU/GPU single computing node and parallel RHSEG using multinode clusters. The multinode clusters includes GPU cluster, hybrid CPU/GPU clusters, CPU clusters and Multicore CPU cluster. The fundamental idea of the solution is the parallelization of the dissimilarity calculation step in the RHSEG algorithm because of the natural suitability of parallelization in these calculations. Other parts of the algorithm are executed on the main CPU thread. The presented work shows that:

- The achieved speedups using single GPU compared to CPU sequential implementation using CUDA platform are 12x, 21.6x using GeForce 550 Ti and Tesla M2030 respectively.

- The achieved speedups using single GPU compared to CPU sequential implementation using C++ AMP platform is 9.6x using GeForce 550 Ti.

- In the hybrid parallel CPU/GPU RHSEG, multicore CPUs were used in cooperation with GPU hardware for the parallel implementation of the RHSEG algorithm. Hybrid RHSEG works by distributing the workload of partitioned quad image sections among different CPU cores which run in parallel and cooperatively with the GPU. For the execution of the RHSEG algorithm on a single GPU and CPU/GPU (8 CPU cores) using a CUDA platform, speedup of 21.6 and 22.8 times sequential CPU is achieved respectively.

- For cluster implementation of the RHSEG algorithm, multi-nodes of both GPU and hybrid CPU/GPU clusters are used. The network cluster is implemented using Amazon Elastic Compute Cloud (EC2), with a number of computing nodes that range from 4 to 16. Cluster RHSEG distributes the partitioned image sections to computing nodes to process them in parallel and collects the results returning them to the main node. For a single node hybrid multicore CPU/GPU and  multi-node computer cluster with 16 nodes for  256x256 image, speedup of 22 and 259 times sequential CPU is achieved respectively.

- The complexity of image, details and number of existing classes don't affect speedups.

- The image depth (number of bands) affect GPU speedups. By increasing the number of bands, the speedups increase and the converse is true.

- Power consumption is reduced to 74% using a single GPU C++ AMP solution compared to equivalent CPU cluster.

- The achievements reported in this work represent a forward step for faster efficient time critical processing for onboard remote sensing.

## 6.2   Future Work

In future, number of optimizations are planned to achieve higher speedups. These optimizations include:

- Significantly optimizing RHSEG using dynamic programming by eliminating the re-computation of unchanged regions in each step.

- Add post processing step to remove the artifacts generated from image splitting

- Changing RHSEG to make multiple region merges per step instead of a single merge per step to reduce the execution time for the first step and reducing the number of steps needed for merging identical regions.

- Tweaking CUDA and C++ AMP GPU implementations by using loop-unrolling techniques and global constant memory for parts of a regions data that are constant during the computation.

Some limitations also will be removed like:

- Limitation on image size. We used square images (N x N images).

- Limitation in GPU implementation for the maximum number of adjacent regions to any region by "max_adjacencies" value. It can be fixed using GPU dynamic arrays for region adjacencies data, thus no longer a limit exists for number of adjacencies for regions.

For CPUs with a high number of cores, more than eight, parallel platforms like OpenMP can be introduced and compared to existing GPU and Hybrid

CPU/GPU parallel implementations. Also for each core, instruction vectorization can be utilized using an enhanced instruction set like streaming SIMD extensions (SSE) for parallel CPU solutions to further achieve higher CPU resources usage. Implementation in more portable GPU and CPU platforms are to be considered like OpenCL and OpenACC.

و يستعرض الفصل الثاني الأبحاث و المحاولات السابقة و المتعلقه بطرق التعرف على الصور فائقة الطيفية وعمل تقسيم للأبحاث والحلول واجراء مقارنة بينهما لإيضاح نقاط القوة والضعف. وينتهي بتوضيح أسباب اختيار خوارزم التعرف الشجري RHSEG للبحث في هذه الرسالة.

أما الفصل الثالث فيستعرض الخلفيه العلمية لخوارزم التعرف الشجري RHSEG المطلوبه لفهم الرسالة. ويشرح الفصل الرابع بالتفصيل الجزء العملي والذي يهدف إلى تطبيق متوازي فائق السرعة لخوارزم "التقسيم الهرمي " RHSEG باستخدام وحدات الرسومات و المعالجات متعددة النوى و الحسابات الموزعة. وذلك باستخدام التقنيات المتوازية CUDA و AMP ++C من شركتي NVidia و Microsoft ، و باستخدام معدات NVidia GeForce و Tesla و عناقيد حوسبة EC2 من شركة.Amazon

أما الفصل الخامس فيشرح بالتفصيل النتائج العملية لمعدلات التسريع واستهلاك الطاقة و دقة التعرف و عقد المقارنة بين النتائج وتأكيد الحصول علي نتائج افضل بما يحقق الهدف من هذه الرسالة. وفي الفصل السادس تم اعطاء الخلاصات الرئيسيةَ المأخوذةَ من هذه الرسالة وكذا نقاط بحث مستقبلية ذات علاقه بموضوع هذه الرسالة.

# ملخص الرسالة

تطبيقات الإستشعار عن بعد في الوقت الحقيقي مثل البحث و الإنقاذ، تتبع الأهداف العسكرية، مراقبة البيئة، منع الكوارث و غيرها من التطبيقات تحتاج إلى إمكانيات معالجة بيانات في الوقت الحقيقي أو القدرة المستقلة على اتخاذ القرارات. عن طريق المعالجة على سطح المركبة يمكن مواجهة تلك التحديات بمعالجة البيانات قبل إرسالها إلى الأرض، فيتم تخفيض حجم المعلومات المرسلة و إعطاء المركبة القدرة على اتخاذ القرار. ولأجل معالجة البيانات على سطح المركبة من المهم توفر معدات معالجة صغيرة الحجم و قليلة استخدام للطاقة. ومع ظهور المستشعرات الضوئية الحديثة الفائقة الطيفية وزيادة أحجام الصور الناتجة منها، وجب أيضا وجود معدات معالجة أكثر قوة. وفي هذا الشأن، تعتبر وحدات المعالجات الرسومية منصة واعدة للمعالجة فائقة الأداء و بحجم صغير.

هدف هذه الرسالة هو بناء برامج معالجة للصور فائقة الطيفية، بحيث تكون فائقة الأداء و عالية الدقة. وبهذا تساهم في بناء أنظمة استشعار عن بعد ذكية و ذات استخدام قليل للطاقة. نقدم في هذه الرسالة تطبيق متوازي فائق السرعة لخوارزم معروف يسمى "التقسيم الهرمي" RHSEG باستخدام وحدات الرسومات و المعالجات متعددة النوى. RHSEG طريقة تم تطويرها من قبل وكالة الفضاء الأمريكية لإعطاء نتائج أفضل عبر مستويات متعددة من المخرجات. تم بناء هذه البرامج باستخدام التقنيات المتوازية CUDA و AMP C++ من شركتي NVidia و Microsoft، و باستخدام معدات NVidia GeForce و Tesla و عناقيد حوسبة EC2 من شركة Amazon. اختبارتنا تبين أن زيادة السرعة من التطبيق المتوازي عن النسخة المتتابعة يصل إلى ٢١ مثلا للوحدة الرسومية الواحدة، و ٢٤٠ مثلا ل ١٦ وحدة مجمعة في عنقود حوسبي. وانخفض استهلاك الطاقة إلى ٧٤٪.

وتعالج الرسالة هذا الموضوع في ستة فصول:

تناول الفصل الأول مقدمة وأسباب اختيار الموضوع و ملخص للمساهمات العلمية للرسالة. كما عرض تمهيد لموضوع البحث وتنظيم فصول الرسالة.

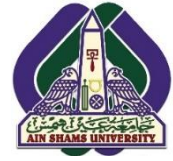

قسم الحسابات العلمية
كلية الحاسبات و المعلومات
جامعة عين شمس

# الحسابات المتوازية و الموزعة لتحليل الصور فائقة الطيفية باستخدام وحدات معالجات الرسوم

رساله مقدمة الى قسم الحسابات العلمية بكلية الحاسبات و المعلومات جامعة عين شمس
كجزء من متطلبات الحصول على درجة الماجستير فى الحاسبات و المعلومات

إعداد

## محمود أحمد حسام الدين محمد

معيد بقسم العلوم الأساسية

كلية الحاسبات والمعلومات

جامعة عين شمس

تحت إشراف

**الأستاذ الدكتور/ محمد فهمي طلبة**

**استاذ بقسم الحسابات العلمية – كلية الحاسبات والمعلومات – جامعة عين شمس**

**د / هالة مشير حسن**

**استاذ مساعد بقسم الحسابات العلمية – كلية الحاسبات والمعلومات – جامعة عين شمس**

**د/ محمد حسن عبد العزيز**

**مدرس بقسم العلوم الاساسية – كلية الحاسبات والمعلومات – جامعة عين شمس**

القاهرة ٢٠١٥